**Towards the optimization of a perovskite-based room temperature ozone sensor: A multifaceted approach in pursuit of sensitivity, stability, and understanding of mechanism**

*Aikaterini Argyrou, Rafaela Maria Giappa, Emmanouil Gagaoudakis, Vasilios Binas, Ioannis Remediakis, Konstantinos Brintakis\*, Athanasia Kostopoulou\*, Emmanuel Stratakis\**

A. Argyrou, E. Gagaoudakis, V. Binas, I. Remediakis, K. Brintakis, A. Kostopoulou, E. Stratakis
Institute of Electronic Structure and Laser
Foundation for Research and Technology-Hellas
Vassilika Vouton, 70013, Heraklion, Greece
E-mail: kbrin@iesl.forth.gr; akosto@iesl.forth.gr; stratak@iesl.forth.gr

A. Argyrou
Department of Chemistry
University of Crete
Vassilika Vouton, 70013, Heraklion, Greece

R. M. Giappa, I. Remediakis
Department of Material Science and Technology
University of Crete
Vassilika Vouton, 70013, Heraklion, Greece

V. Binas
Department of Chemistry
Aristotle University of Thessaloniki
Thessaloniki, 54633, Greece

Keywords: mixed halide perovskites, Mn-doped mixed halides, halide vacancies, perovskite long-term stability, perovskite degradation, adsorption energy, density-functional theory

Metal halide perovskites (MHPs) have attracted significant attention owing to their simple manufacturing process and unique optoelectronic properties. Their reversible electrical or optical properties changes in response to oxidizing or reducing environments make them prospective materials for gas detection technologies. Despite advancements in perovskite-based sensor research, the mechanisms behind perovskite-gas interactions, vital for sensor performance, are still unexclusive. This work presents the first evaluation of the sensing performance and long-term stability of MHPs, considering factors such as halide composition variation and Mn doping levels. The research reveals a clear correlation between halide composition and sensing behavior, with Br-rich sensors displaying a p-type response to $O_3$ gas, while Cl-based counterparts exhibit an n-type sensing behavior. Notably, Mn-doping significantly enhances the $O_3$ sensing performance by facilitating the gas adsorption process, as supported by both atomistic simulations and experimental evidence. Long-term evaluation of the sensors provides valuable insights into evolving sensing behaviors, highlighting the impact of dynamic instabilities over time. Overall, this research offers insights into optimal halide combination and Mn-doping levels, representing a significant step forward in engineering room temperature perovskite-based gas sensors that are not only low-cost and high-performing but also durable, marking a new era in sensor technology.

## 1. Introduction

During the past few decades, the increased industrial activity and rapid urban growth have given rise to the emission of toxic, harmful and explosive gases, contributing significantly to air pollution and posing substantial risks to both the environmental and human safety.[1,2] As a consequence, the development of robust gas sensors, capable of detecting and monitoring ultra-low concentrations of those hazardous compounds, is essential. Among a wide range of semiconducting gas sensing elements, metal oxide-based sensors were the first reported and remain the most thoroughly studied.[3] The evolution of metal oxide gas sensors over the years has been driven by continuous research, which, among others, has included the influence of morphology, particle size and doping in gas-metal oxide interactions.[4–6] Recent studies have also explored stability issues related to long-term conductivity and response, leading to the implementation of improvement strategies that have contributed to recent advances in metal oxide gas sensors.[7,8] However, despite the extensive research, metal oxide-based sensors operate at high temperatures, increasing the power consumption, the overall device size and cost of gas sensors.[9] Conversely, room temperature metal oxide based sensors face challenges related to insufficient sensitivities, prolonged response/recovery times and poor reversibility.[10] Therefore, there is a high demand for new materials that could overcome the former limitations. Recently, metal halide perovskites (MHPs) with the chemical formula $ABX_3$, where A is an organic or inorganic cation, B is a divalent metal and X is a halide anion, have been reported as potential gas sensing elements for the detection of a wide range of gaseous pollutants, with very promising results in terms of responsiveness and selectivity.[11–14] These sensors operate efficiently at room temperature, offering an advantage over the traditional metal oxide gas sensors, eliminating additional energy consumption and enabling the development of portable gas sensing devices.[15,16] The interactions between MHPs and gas molecules, which characteristically encompass charge transfer, gas-induced defects, and defect passivation,[17–20] are inherently complex and multifaceted. Owing to this complexity, these interactions have not yet been thoroughly investigated and remain a subject of ongoing scientific inquiry.

Despite the advancements in the detection of various target gases, the field of MHP gas sensors requires further exploration into material design facets and pivotal parameters, including the impact of material composition and aging on operational efficacy. In-depth investigations in these areas and on the underlying sensing mechanism could significantly inform the design of materials, thereby enhancing the efficiency of MHP gas sensors.

While the challenges and potential of MHPs for gas sensing are under exploration, one particular area of interest within the realm of gas sensors is the detection of $O_3$ pollutant, a highly reactive and oxidizing agent which is widely used in medicine and agriculture.[21,22] Ozone is a critical component of the Earth's atmosphere, playing a vital role in absorbing harmful ultraviolet radiation. However, at ground level, ozone is a major air pollutant with detrimental effects on human health, vegetation, and materials.[23,24] In recent years, an increase in the concentration of ozone gas has been observed, particularly in densely populated areas highlighting the urgent need for sensitive ozone sensors for air-quality monitoring, assessing the effectiveness of pollution control measures, and ensuring public safety.[25]

Conventional ozone sensing technologies often rely on expensive and complex instrumentation, such as ultraviolet photometric detectors or chemiluminescence analyzers. These methods can be costly to implement and maintain, limiting their widespread deployment, particularly in resource-constrained settings. Additionally, some existing sensors may suffer from cross-sensitivity to other gases, leading to inaccurate readings.[26]

In response to the need for efficient ozone detection, MHPs, both hybrid organic-inorganic and all-inorganic, have been investigated as prospective materials. Their enhanced sensitivity and capability to operate at ambient temperatures facilitate the development of compact and lightweight sensing devises. $CH_3NH_3Pb_{3-x}Cl_x$ thin film was initially reported as an effective ozone sensing element, however, concerns arising from their long-term stability and potential susceptibility upon $O_3$ exposure remain unresolved.[27] Contrastingly, our research group has reported the development of ultrasensitive, ligand-free $CsPbBr_3$ microcrystals characterized by markedly improved stability. Additionally, even slight morphological alterations among these microcrystals have been observed to manifest significant differences in $O_3$ detection responses.[28,29] This phenomenon was attributed to the presence of surface defects, providing insightful contributions toward clarifying the underlying sensing mechanisms.

Considering that the charge transport in halide perovskites is predominantly facilitated by the interaction between lead and halide atoms, which significantly influence the conduction band maximum (CBM) and valence band minimum (VBM),[30,31] a detailed investigation into the interaction of $O_3$ gas with metal cations and halide anions is crucial. Such a study is expected to offer valuable insights into the fundamental sensing mechanisms. In this context, the present research is centered on examining the sensing properties of both metal-undoped and Mn-doped $CsPbBr_{3-x}Cl_x$ mixed halide perovskite microcrystals (µCs). The strategic incorporation of Mn as a dopant is based on its recognized role in altering the electronic properties of halide perovskites.[32] Furthermore, Mn is known for its catalytic activity in decomposing $O_3$, suggesting its potential to enhance sensor performance.[33] It is established, that partial substitution of Pb with Mn within the $CsPbBr_{3-x}Cl_x$ framework leads to a narrowing of the bandgap, thereby facilitating electron mobility from the valence band to the conduction band and improving both light absorption and electrical conductivity.[34] Mn-doped perovskites have also been successfully utilized as oxygen-responsive optical probes.[35] In more details, pre-synthesized $CsPbBr_3$ µCs are modified to introduce Mn-doping via anion exchange. Additionally, reference samples of $CsPbBr_3$ and $CsPbCl_3$ µCs, maintaining consistent morphology, are prepared for comparative analysis. A series of sensing experiments in response to $O_3$ gas exposure, complemented by atomistic simulations elucidate the preferential sites for the gas adsorption in both doped and undoped µCs by tuning the halide ratio and Mn-doping level. This investigation contributes to understand the influence of halide anions and metal cations on the electrical conductivity and $O_3$ sensitivity of MHPs. The study further monitors the temporal progression of the sensing performance, simultaneously assessing the changes in the optical, structural, and chemical properties of the sensing materials. This investigation further sheds light on the degradation pathways of all-inorganic MHPs, thereby contributing to a deeper understanding of their long-term stability and durability.

## 2. Results and Discussion

### 2.1 Sensing elements' design: synthesis and features

Ligand-free, metal-undoped and Mn-doped mixed halide perovskites ($CsPbBr_{3-x}Cl_x$, 0<x<3) in the form of µCs with controlled anion ratio and Mn-doping level were prepared by modifying pre-synthesized $CsPbBr_3$ μCs via room temperature cation and anion exchange processes. In the case of undoped mixed halide μCs, this was achieved through an anion exchange process, by introducing precise volume-to-volume ratios of a chloride-containing precursor ($PbCl_2$) into the solution containing the $CsPbBr_3$ μCs (20%, 50% and 80% v/v). For the Mn-doped systems, molecular doping was performed using the same volume-to-volume ratios of a $MnCl_2$ precursor, enabling controlled Mn doping. The first method followed a conventional anion exchange approach for synthesizing mixed halide nanocrystals, while the second method employed a Mn-doping approach based on a halide exchange-driven cation exchange (HEDCE) technique, as previously reported by Hung et al. and originally applied for the synthesis of Mn-doped $CsPb(Cl/Br)_3$ nanocrystals.[36,37] For comparison, $CsPbBr_3$ (0% v/v) and $CsPbCl_3$ (100% v/v) μCs were synthesized via a room temperature re-precipitation-based method by introducing a small amount of the precursor solution ($CsBr/CsCl$ and $PbBr_2/PbCl_2$ in DMF) into the bad solvent, namely toluene (§4. Experimental Section/Methods).

*2.1.1. Undoped all inorganic mixed halide perovskite μCs*

The undoped mixed halide $CsPbBr_{3-x}Cl_x$ perovskite systems exhibited a cubic-shaped morphology as observed by scanning electron microscopy (SEM) images with an average size of about $0.95 \pm 0.12$ μm, (Figure 1a and S1a-c, Supporting Information). These morphological features were similar to those of the pre-synthesized $CsPbBr_3$ μCs (0% v/v) used in the anion exchange process (Figure S1d, Supporting Information), as well as those of the $CsPbCl_3$ μCs (100% v/v) synthesized for comparison (Figure S1e, Supporting Information). A reduction of the microcrystal size occurred during the anion exchange process, which could be attributed to the different sizes between Br and Cl anions (Figure S1f, Supporting Information).

The elemental composition of the synthesized materials was estimated by Energy Dispersive Spectroscopy (EDS), as shown in Figure S2, and summarized on Table S1 (Supporting Information). The chemical composition of the preformed $CsPbBr_3$ μCs was verified by X-ray Photoelectron Spectroscopy (XPS), corroborating the EDS results (Figure S2a and S3, Supporting Information). Particularly, the % atomic concentration, calculated from the peak areas of Cs 3d, Pb 4f and Br 3d of the XPS survey spectrum, was determined to be 22.47% Cs, 19.4% Pb and 58.12% Br, with relative atomic concentration of Cs:Pb:Br=1.2:1:3.1 (Figure S3, Supporting Information). Moreover, the Bragg reflections observed in the XRD pattern of the $CsPbBr_3$ μCs matched well to the orthorhombic crystal structure of $CsPbBr_3$ (ICDD 01-085-6500), and no traces of secondary phases or impurities were detected (Figure 1b, black pattern). Moreover, a shift of the diffraction peaks towards higher angles was observed in the $CsPbBr_3$- $_x$Cl$_x$ μCs, indicating the gradual substitution of Br$^-$ ions with Cl$^-$ ions in the crystal lattice and the successful anion exchange (Figure 1b, red, blue and magenta patterns). This gradual substitution of Br$^-$ ions with Cl$^-$ ions in the crystal lattice led to its contraction which is associated with the different bond length between Pb-Br and Pb-Cl.[30] Similar shifts of the diffraction peaks to higher angles, indicative of a cell shrinkage have been reported for CsPbBr$_3$ nanocrystals upon the incorporation of Cl$^-$.[32,38,39] Additionally, as the Cl concentration increases, it is noticed that the peaks between 24º to 30º gradually diminish, leading to the transformation from orthorhombic to tetragonal CsPbCl$_3$ structure (Figure S4, Supporting Information). It is important to note here, that the diffraction peaks of the CsPbCl$_3$ μCs matched with the tetragonal CsPbCl$_3$ phase (ICSD 18-0366), with small traces of the monoclinic Cs$_4$PbCl$_6$ phase (ICSD 78-1207) (Figure 1b, green pattern).

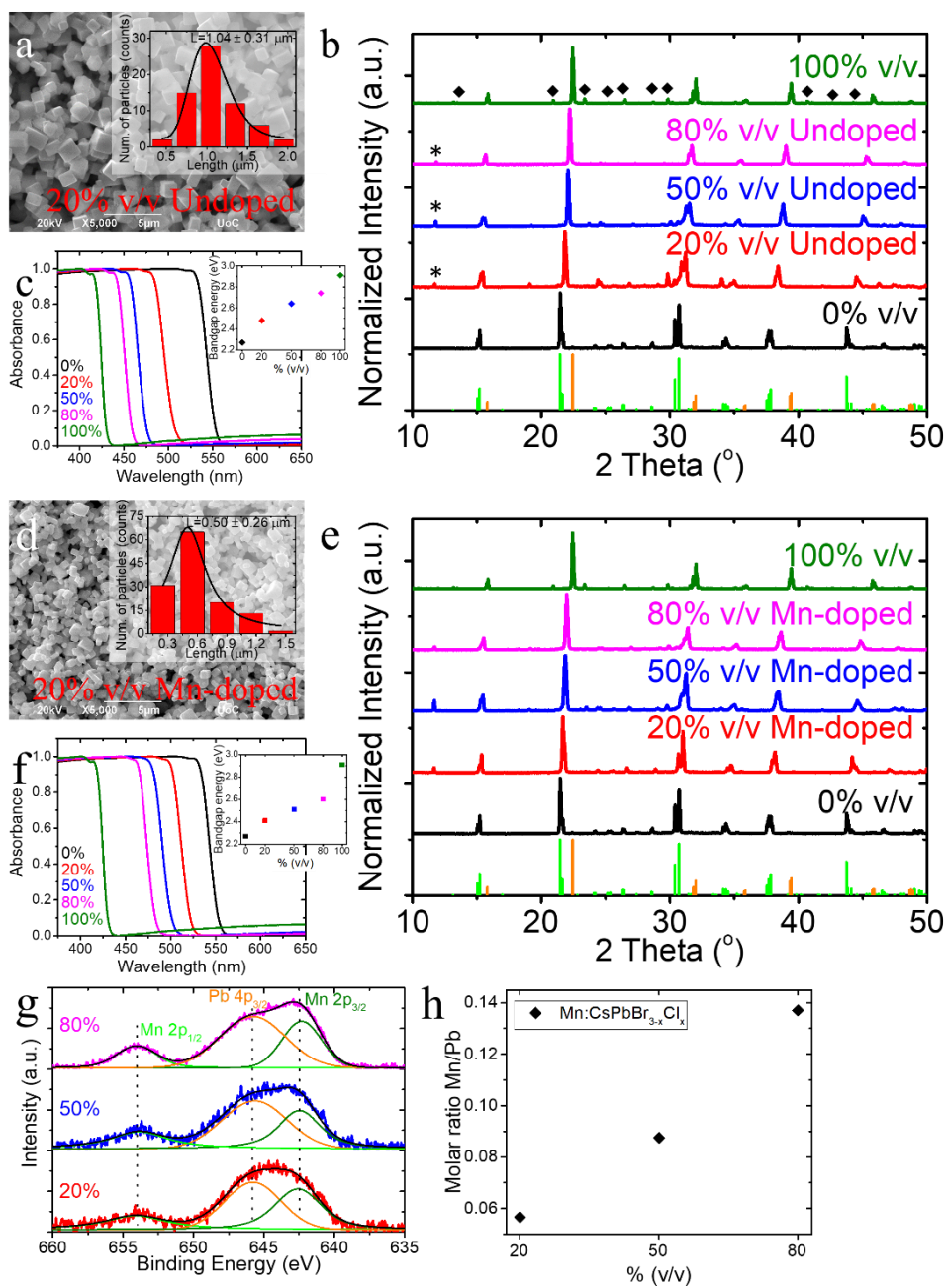

**Figure 1**. a) SEM image of 20% v/v undoped CsPbBr$_{3-x}$Cl$_x$ μCs. The inset showed the size distribution diagram of the corresponding sample. b). XRD patterns from undoped halides with different v/v ratio and the reference samples, CsPbBr$_3$ (0% v/v), and CsPbCl$_3$ (100% v/v). XRD reference patterns of the orthorhombic CsPbBr$_3$ crystal structure (ICDD 01-085-6500, light green pattern), and the tetragonal CsPbCl$_3$ (ICSD#18-0366, orange pattern) were included for comparison. The asterisks and rhombus denoted the tetragonal CsPb$_2$Br$_5$ (ICSD 00-025-0211) and monoclinic Cs$_4$PbCl$_6$ phase (ICSD#78-1207), respectively. c) Absorbance spectra and bandgap energy as calculated by Tauc plot method (inset) for the same CsPbBr$_{3-x}$Cl$_x$ samples. d) SEM image of 20% v/v Mn-doped mixed halide μCs and corresponding inset represented

the μCs' size distribution diagram. e) XRD patterns of the mixed halide μCs with different volume-to-volume ratio and the reference $CsPbBr_3$ (0% v/v). XRD reference patterns of the orthorhombic $CsPbBr_3$ crystal structure (ICDD 01-085-6500, light green pattern), and the tetragonal $CsPbCl_3$ (ICSD#18-0366, orange pattern) were included as well. f) Absorbance spectra and bandgap energy as calculated by Tauc plot method (inset) for the same samples. (g) High resolution XPS spectra of the same samples. The deconvolution of Mn 2p revealed the presence of Mn 2p3/2 (green line), Pb 4p3/2 (orange line) and Mn 2p1/2 (light green line). h) Mn-to-Pb molar ratios of Mn-doped systems as determined by ICP-MS.

The anion exchange was further demonstrated by the high-resolution Br 3d and Cl 2p core level XPS spectra (Figure S5, Supporting Information). The binding energy of Br 3d of mixed-halide systems exhibited a shift of approximately ~0.3 eV upon the introduction of Cl atoms in the $CsPbBr_3$ crystal lattice, which can be assigned to the electronegativity difference between Br and Cl atoms that could affect the binding energy of Br 3d core level (Figure S5, Supporting Information). It was also noticed that the relative intensities between Br 3d and Cs 4d decrease with the increase of Cl content, suggesting further the successful substitution of $Br^-$ with $Cl^-$. Furthermore, a blue-shift in the binding energy of the Pb 4f doublets with spin-orbit splitting energy 4.9 eV was observed, indicating modifications in the chemical environment on the surface of the materials. This shift can be assigned to the partial substitution of $Br^-$ ions by $Cl^-$ ions within the $Cs-(PbX_6)$ octahedral (Figure S6, Supporting Information).[38,40]

The successful anion exchange in the $CsPbBr_3$ μCs was also confirmed by UV/Visible (UV/Vis) spectroscopy, revealing band gap energy tuning as a function of the Cl-content (Figure 1c). The absorbance spectra exhibited a clear blue-shift of the absorption band edge with increasing $Cl^-$ content, indicating an increase in the band gap energy, which was calculated using Tauc plot method (Figure 1c inset). The band gap energy was found to increase from 2.27 eV to 2.91 eV as the volume-to-volume (v/v %) ratio increased from 0 to 100%.

*2.1.2 Mn-doped all inorganic mixed halide perovskite μCs*

The Mn-doped $CsPbBr_{3-x}Cl_x$ μCs were synthesized by a modified halide exchange-driven cation exchange (HEDCE) approach.[37] According to this method, the use of $MnCl_2$ molecules is essential for achieving simultaneous halide and cation exchange within the same lattice site. The cubic-shaped morphology of the μCs remained unchanged during the cation exchange process, even after increasing the volume of the Mn-containing precursor (Figure 1d and S7, Supporting Information). However, the μC size was reduced from 0.95 ± 0.12 to 0.49 ± 0.02 μm (Figure S1 and S7, Supporting Information), while remaining almost unchanged with further increase of the Mn-precursor volume. The initial size reduction could be attributed to the different sizes between Br and Cl anions, as well as the Pb and the inserted Mn cations, while the constant size could be due to the small percentage of the Mn-inserted cations. The elemental composition was estimated by EDS (Figure S8 and Table S2, Supporting Information) while the cation and anion exchange upon the addition of the $MnCl_2$ precursor were confirmed by the XRD patterns and the XPS spectra. Similar to the undoped mixed halide systems, the XRD patterns of Mn-doped systems demonstrated a monotonic shift towards higher angles with increasing the amount of $MnCl_2$ precursor (Figure 1e). This shift was attributed to both the substitution of $Br^-$ with $Cl^-$ anions and $Pb^{2+}$ with $Mn^{2+}$ anions. Furthermore, similar to the undoped systems, Mn-doped $CsPbBr_{3-x}Cl_x$ exhibited comparable XPS spectral characteristics attributed to anion exchange phenomena (Figure S9 and S10, Supporting Information). The gradual increase in $Mn^{2+}$ concentration in the $CsPbBr_{3-x}Cl_x$ μCs upon increasing the volume percentage was also confirmed by the enhanced relative intensity of Mn $2p_{3/2}$ peak compared to Pb $4p_{3/2}$ peak in high resolution XPS spectra (Figure 1g). Nevertheless, the molar concentration of the Mn in the Mn-doped μCs was determined by ICP-MS as the quantification of $Mn^{2+}$ from the XPS survey spectra proved challenging due to the complexity of deconvoluting the Mn 2p peaks, given the overlapping of Pb $4p_{3/2}$ peak interference (Figure 1h).

Similar to the undoped mixed halide systems, the absorbance spectra of the Mn-doped sensors showed a blue-shift of the band edge as the volume-to-volume ratio was increased, resulting in the band gap energy increase (Figure 1f). This raise was originated from the synergetic combination of the of Cl-to-Br molar ratio increase in the $CsPbBr_3$ host lattice but also to the incorporation of $Mn^{2+}$. The latter contribution, according to the literature in similar systems, causes slight blue shifts of the absorbance spectra and may be attributed to alloying effects on the perovskite band structure.[41]

Moreover, despite maintaining consistent ratios of $PbCl_2$ and $MnCl_2$ precursor solutions at 20%, 50% and 80% v/v for the fabrication of undoped and Mn-doped $CsPbBr_{3-x}Cl_x$ μCs respectively, as well as, the same reaction times, variations in halide content were observed (Figure S11a, Supporting Information). The Br content decreased as the volume-to-volume ratio increased, and conversely, an increase was observed for the Cl content for both undoped and Mn-doped μCs. However, the Cl-content was lower in the Mn-doped systems across all volume-to-volume ratios. These variations may be attributed to the simultaneous anion and cation exchange and the choice of $MnCl_2$ as an anion exchange precursor. The $[PbBr_6]^{4-}$ octahedral open up via extended halide exchange including also the partial substitution of Pb sites. This process occurs at longer reaction times than in the undoped systems due to the difficulty of the molecule $MnCl_2$ to diffuse into the crystal lattice.[37,32] This finding was further supported by comparing the absorbance spectra and the band gap energies of the undoped and Mn-doped systems (Figure S11b-c, Supporting Information). Contrary to the anticipated larger blue shifts and higher bandgap energies for the Mn-doped systems, the opposite was observed suggesting a slower diffusion of $MnCl_2$ compared to $PbCl_2$.

## 2.2 Ozone sensing performance of all-inorganic metal halide perovskite µCs

The precipitated perovskite µCs were deposited onto commercial interdigitated platinum electrodes supported on glass substrate and left to dry under vacuum. Subsequently, the ozone detection capability of all fabricated perovskite-based sensors was assessed through room-temperature electrical measurements under total dark operating conditions. The gas sensing experiments were carried out in a home-made chamber with a constant pressure of 700 mbar whereas a voltage of 2 V to 3.5 V was applied. The sensing process was initiated by exposing the sensors to ozone gas for 150 s, followed by a 200 s recovery with synthetic air. All samples were exposed to different ozone concentrations ranging from 1567 down to 6 ppb (§4. Experimental Section/Methods).

### 2.2.1 Ozone sensing performance of mixed halide perovskites

The electrical response diagrams of each sensor upon $O_3$ exposure revealed the significant influence of the halide composition in the sensing process (Figure 2). Notably, the undoped 50% and 80% v/v based-sensors exhibited a decrease in current intensity upon $O_3$ exposure, followed by an increase upon synthetic air (Figure 2 c-d). This trend was observed for all tested $O_3$ concentrations, ranging from 1567 ppb down to 4 ppb, indicating an n-type sensing behavior, akin to that observed in $CsPbCl_3$ µCs (Figure 2e). On the other hand, the 0% v/v $CsPbBr_3$ based-sensor displayed a contrasting response pattern, and particularly, the current was increased till a maximum current value upon $O_3$ exposure, followed by a recovery to its baseline in the presence of synthetic air, suggesting a p-type behavior (Figure 2a), as has been previously shown by our group.[29,28] By correlating these results with the halide content diagram obtained from the XPS spectra (Figure 2f inset and S12, Supporting Information), it was observed that the sensors with higher Br content (Br-rich) displayed a p-type behavior, while the Cl-rich ones exhibited n-type sensing behavior. This conclusion was further supported by the sensing behavior of the Br-rich 10 % v/v sensor, which found to be analogous to the $CsPbBr_3$-based sensor (Figure S13, Supporting Information). It is important to note here that the nearly equimolar, regarding the halide content, the 20% v/v perovskite-based sensor was unable to detect $O_3$ at any concentration, as the competing p- and n-type sensing behaviors intersected, resulting in the absence of $O_3$ sensing behavior (Figure 2b).

Moreover, the response calculated for the highest $O_3$ concentration showed that the mixed halide perovskites-based sensors maintain a nearly constant response as the volume-to-volume ratio increases, which was similar to that of the $CsPbCl_3$ sensor, but considerably lower than the $CsPbBr_3$ sensor. This observation indicated that sensors with higher Br content exhibited higher response, which decreased with the incorporation of $Cl^-$ ions within the crystal lattice (Figure 2f).

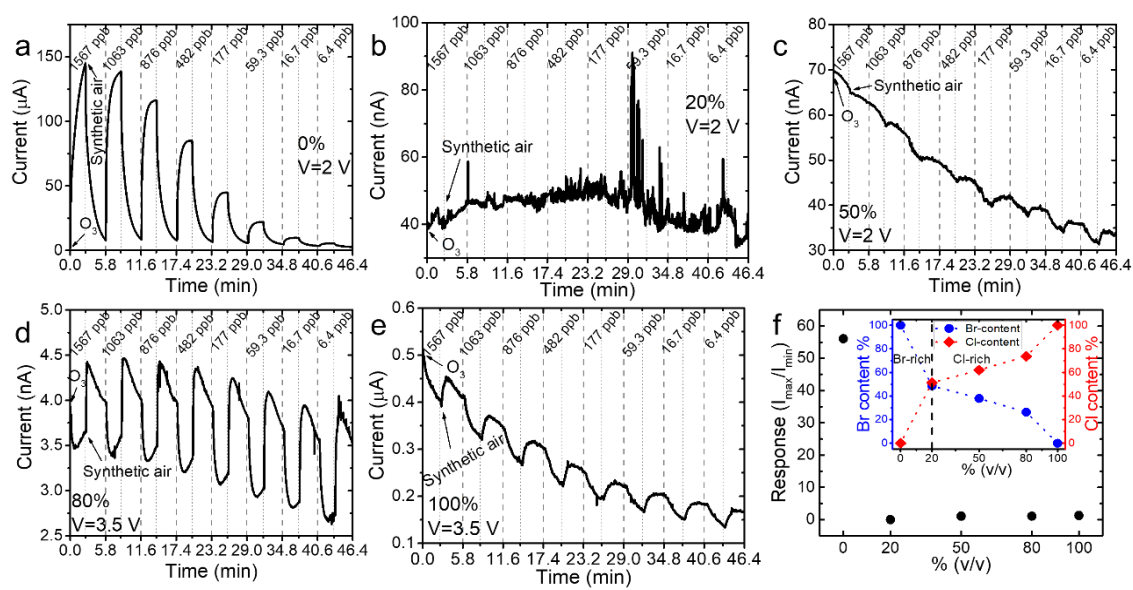

**Figure 2.** $O_3$ sensing performance of the undoped perovskite-based sensors, by varying the volume-to-volume ratio from 0 to 100% (a-e). (f) Calculated response of the undoped sensing systems at 1567 ppb and as inset the halide content of the undoped μCs as calculated by XPS spectra.

*2.2.2 Ozone sensing performance of Mn-doped mixed halide perovskites*

Figure 3 illustrates the performance of Mn-doped $CsPbBr_{3-x}Cl_x$ perovskite-based sensors upon $O_3$ exposure. Intriguingly, the 20% and 50% v/v sensors demonstrated a p-type behavior in response to $O_3$ gas (Figure 3a-b), while the 80% v/v Mn-doped sensor showed no sensing capability (Figure 3c), contradicting the expected n-type behavior associated with its higher concentration of Cl relative to Br ions. This observation suggests a potential overlap in the roles of Mn and Br ions with the influence of Cl ions in the overall sensing behavior. These results contrast with the corresponding undoped mixed halide sensors, which exhibited an n-type behavior for the 50% and 80% volume-to-volume ratios, and no sensing capability for the 20% v/v sensor (Figure 2b-d). Notably, the Mn-doped perovskite-based sensors showed an improved response at 1567 ppb compared to their undoped counterparts, with the maximum improvement for the 20% v/v Mn-doped sensor (Figure 3d). This sensor also displayed the highest response among all Mn-doped sensors across the entire range of $O_3$ concentrations (Figure 3e). These results indicated that Mn-doping has a significant and specific role on the sensing behavior, but it was not easily distinguished from that of the halide contribution. It is essential to recall here what was mentioned previously, that the sensors of the same volume-to-volume ratio do not have the same Br to Cl ratio (Figure 2d inset and S14, Supporting Information).

To assess the impact of Mn-doping on sensing performance, a comparison in the sensing behavior between Mn-doped and undoped systems with similar Br-to-Cl ratio was essential. Notably, both 20% v/v undoped and 50% Mn-doped-based sensors demonstrated nearly equimolar halide content, however significant variations in the sensing performance were observed (Figure 3d inset). Particularly, the 20% v/v undoped sensor did not exhibit any sensing capability, as previously discussed (Figure 2b) while the 50% Mn-doped based-sensor showed a p-type sensing behavior (Figure 3b), suggesting that Mn-doping can override the synergetic effect of the halides. Moreover, the influence of Mn on the sensing performance was evident in the 50% v/v undoped and 80% v/v Mn-doped based systems demonstrate equimolar halide content as well. The strong effect of Mn-doping was further studied by comparing the response between the 10% undoped (Figure S13, Supporting Information) and the 20% Mn-doped sensors (Figure 3a). Given that both sensors are characterized by higher concentration of Br compared to Cl ions, a p-type behavior was observed, as expected (Figure 3a and S13, Supporting Information). However, while both systems were capable of detecting low ozone concentrations, the Mn-doped system displayed a capability to effectively detect ultra-low gas concentrations below of 15 ppb (Figure 3a) in contrast to the undoped one (Figure S13, Supporting Information). The calculated response of the 20% v/v Mn-doped system surpassed that of the undoped counterpart by more than three times (Figure 3f).

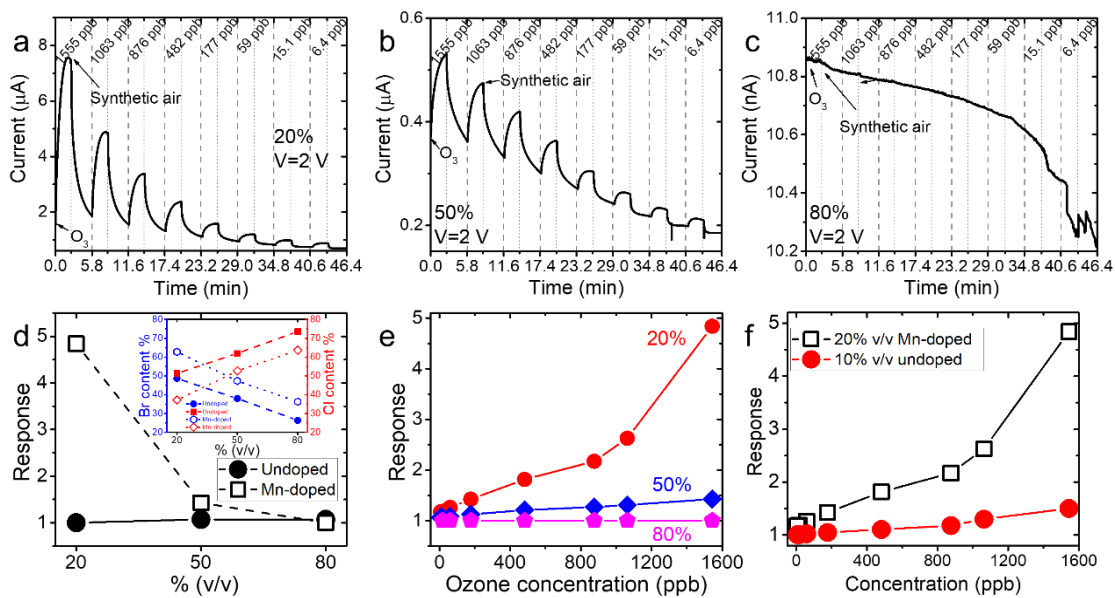

**Figure 3.** O$_3$ sensing performance of the Mn-doped a) 20%, b) 50% and c) 80% v/v perovskite-based sensors. d) Calculated response of Mn-doped (white circle) and undoped (black circle) sensing systems at 1567 ppb O$_3$ concentration. Halide content of the Mn-doped μCs calculated by XPS spectra (inset). e) Response of the Mn-doped sensors varying the Mn-doping level as a function of O$_3$ concentration. f) Response comparison diagram of 10% v/v undoped (red circle) and 20% v/v Mn-doped sensor (white circle) as function of O$_3$ concentration.

## 2.3. Ozone sensing in mixed halide perovskites through site-specific analysis and computational modeling

A thorough understanding of the ozone sensing mechanism in MHPs requires identifying the specific sites where O$_3$ molecules interact with the perovskite surface. Previous studies have identified halide vacancies and trap states as critical sites for gas adsorption/desorption, a phenomenon substantiated by observable alterations in the material's electrical conductivity. However, it is noteworthy that only a singular study examining the sensing behavior of a perovskite film in the presence of oxygen has corroborated these findings with theoretical calculations.[19,27,42,43]

In general, a good sensing material could offer strong binding to the target gas while it could greatly alter its electronic structure upon adsorption. Both these properties can be probed using DFT simulations, which can provide accurate adsorption energies, which is a measure of the strength of the gas-solid interaction, and density of states, which offers insights on the change of electronic properties of the material during sensing (§S2, Supporting Information). To elucidate the sensing properties of these perovskite materials, DFT simulations have been employed for a series of model systems. Simplified surface models and $O_2$ molecules instead of $O_3$ have been used for computational efficiency, as prior DFT investigations focused on $O_3$ sensors, have indicated that adsorbed $O_3$ predominantly appears as either oxygen molecules or atoms. The binding of $O_2$ and $O_3$ to polar surfaces relies on electron charge transfer toward surface atoms and the coupling of oxygen's s and p orbitals with the s and p orbitals of an electropositive surface atom.[44–46]

The (001) surface of the $CsPbBr_3$ has been utilized as the model surface, owing to its demonstrated stability, corroborated previously by the findings of Xing et al.[42] Moreover, since halide- and metal-atom vacancies can introduce trap states and act as active sites for gas molecule absorption/desorption upon $O_3$ exposure, vacancies have been introduced in the models.

The effect of anion exchange and cation doping, as well as the formation of defects were investigated using DFT by considering different 5-layer surface models (Figure 4a-e); (a) without defects, (b) with Br vacancy in the outmost surface layer, (c) with Pb vacancy in the outmost surface layer, (d) with Mn doping, where 1/4 of the outmost surface Pb atoms are substituted by Mn and (e) with Cl doping, where 1/8 of the outmost surface atoms are substituted by Cl. For case (e), systems with higher Cl concentrations as well as systems with Cl vacancies (see additional models in §S2.1, Supporting Information).

The calculated absorption energies of the above models (Figure 4f and Table S3, Supporting Information) showed that the $O_2$ adsorption was energetically more favorable at Br-vacancies ($\Delta E_{ads}$= -1.69 eV) compared to the defect-free surface ($\Delta E_{ads}$= 0.71 eV) and Cl-doped surface ($\Delta E_{ads}$= 0.73 eV). This indicated that oxygen binding was expected to occur primarily close to halide vacancies, as the adsorption energies for the defect-free case suggested minimal interaction between molecular oxygen and the surface. This absence of interaction was further observed for the Cl doped surface with all Br atoms in the uppermost layer replaced by Cl atoms, as well as the $CsPbCl_3$ surface ($\Delta E_{ads}$=0.75 eV and 0.76 eV, respectively), (Figure S15b and c, Supporting Information). However, similar to the $CsPbBr_3$ surface with Br-vacancy, molecular oxygen was bound strongly in the Cl-vacancy site of the defected $CsPbCl_3$ surface ($\Delta E_{ads}$=-1.87 eV) (Figure S15d, Supporting Information). It should be noted, the positive binding energy of the oxygen molecule on defect-free perovskite surfaces was a result of the change in the electron configuration of the molecule upon adsorption from paramagnetic to diamagnetic.

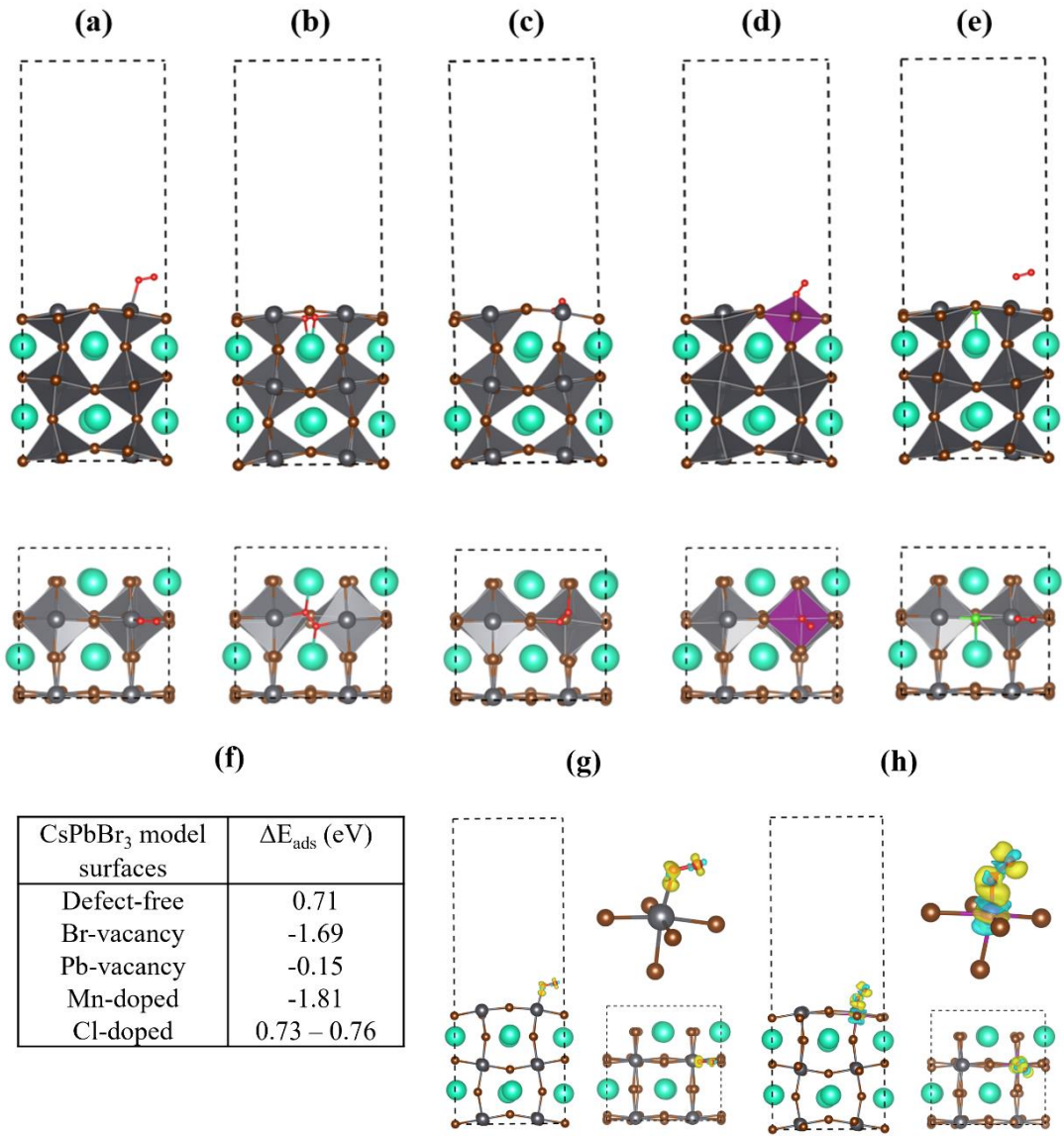

**Figure 4.** DFT-optimized configurations of the model surfaces viewed along b (upper line) and c (bottom line) axis; a) defect-free CsPbBr$_3$, b) CsPbBr$_3$ with Br vacancy and c) with Pb vacancy, d) Mn-doped CsPbBr$_3$ slabs and e) Cl-doped. Cs, Pb, Br, Mn, Cl, and O atoms are illustrated as cyan, gray, brown, magenta, green, and red atoms, respectively. Dashed lines indicate the periodic boundaries of the simulation supercell. f) Adsorption energies of O$_2$ adsorbed on the defect-free (001) surface of CsPbBr$_3$, at the Br and Pb vacancies, and at the Mn- and Cl- doped modified surfaces. The range of values for the Cl-doped system corresponds to different Cl concentrations examined. g-h) Charge density difference plots calculated at 0.005 (e/bohr$^3$) isosurface for: g) O$_2$ adsorbed on the defect-free and h) Mn-doped CsPbBr$_3$ surface, viewed along the b (side view) and c (top view) axis, and magnified at the corresponding adsorption sites at the upper right plots. Yellow and cyan areas illustrate charge accumulation and depletion, respectively.

Similar to Br-vacancy, Mn sites in the Mn-doped surfaces were bound strongly to molecular oxygen, exhibiting an even greater enhancement of the binding energy ($\Delta E_{ads}$= -1.81 eV), contributing further to the gas sensing response enhancement. This is a remarkable finding, indicating that Mn doping alone could introduce oxygen binding sites comparable in strength to halide vacancies. To better understand this enhancement, charge density difference plots were calculated for the defect-free and the Mn-doped surfaces, visualizing the changes in the charge density of these surfaces upon interaction with oxygen (Figure 4g-h). As it could be clearly seen, Mn-doped surface exhibited stronger polarization of the oxygen electron cloud and a more prominent polarization of the charge density of the Mn atoms. This stronger polarization effect of Mn compared to Pb followed the fact that Mn atoms are less electronegative than Pb, thus having a larger electronegativity difference with oxygen atoms. Accordingly, the Mn-O bond is more polar than Pb-O, a fact also reflected in bond lengths, with $d_{Mn-O}$ and $d_{Pb-O}$ being 1.71 and 2.61 Å, respectively. The introduction of specific impurities, either by adding electron-rich or electron-deficient sites in a semiconductor modifies its band structure by introducing energy levels associated with the dopants.

To gain a more comprehensive understanding on the impact of the different doping scenarios on the electrical conductivity and $O_3$ sensing behavior, the electronic band structure (BS) and the partial density of states (PDOS) were calculated for all model cases under study (Figures S16-S22, §S2.1 and S2.2 all the details for the calculations, Supporting Information). Minimal electronic structural changes were caused by bromide substitution with chlorine, maintaining the defect-free surface's characteristics, as chlorine's states lie deep within the band structure. However, increasing chlorine substitution amplifies its valence band contribution, nudging the Fermi level towards n-type doping characteristics. On the other side, new donor states in the conduction band and localized d orbital states within the bandgap were introduced by manganese doping. Upon adsorption of oxygen on Mn-doped model surface, the shallow donor states were quite withdrawn, and new acceptor states appeared at the top of the valence band. The d-states of Mn remained localized at the same relative position within the gap, while the oxygen states exhibit a non-negligible density located both at the valence band and close to the conduction band minimum.

It should be noted that calculation of fine details of the electronic structure such as the precise position of impurity levels should be handled with caution as it could be affected by the employed models. For example, slab-based simulations mimic surface doping and not bulk doping; the former can result in localized electronic states on the surface that can alter the electronic structure.[47] On the other hand, DFT simulations effectively explain both the strong effect of Mn doping on adsorption energy and the changes in the electronic structure of all systems upon oxygen adsorption, as in all cases oxygen alters dramatically the electronic states.

## 2.4. Evaluation of the sensing capability and material stability of lead halide perovskite based-sensors over time

The sensing capabilities of each sensor were re-evaluated after a month of storage under ambient conditions, referred to as aged sensors. Interestingly, both undoped and Mn-doped mixed halide perovskite-based sensors, as well as the pure $CsPbBr_3$ and $CsPbCl_3$ μCs exhibited a conductivity improvement over time (Figure S23-25, Supporting Information). Notably, the undoped 20% v/v-based sensor showed a slight enhancement upon $O_3$ exposure over time, compared to its initial absence in sensing signal (Figure S24a, Supporting Information), while the most prominent improvement upon time was observed in the 80% v/v Mn-doped sensor, which initially displayed no sensing capability (Figure S24c, Supporting Information). Additionally, the sensing behavior (p- or n-type) was preserved after storage under ambient conditions for all the sensors, except for the 50% v/v undoped mixed-halide sensor fabricated, which demonstrated an n-type to p-type conductivity transition upon $O_3$ exposure over time (Figure S24b, Supporting Information).

Furthermore, a comparison between the initial response and that of the aged undoped sensors for various $O_3$ concentrations revealed that all systems displayed a fairly stable response over time (Figure S27, Supporting Information), with a percentage response difference close to zero (Figure 5a, equation 2). This behavior is analogous to that observed in 100% v/v $CsPbCl_3$ sensor (Figure 5a and S26b, Supporting Information).

On the other hand, Mn-doped sensors showed instabilities regarding their response over time. Particularly, the 20% v/v Mn-doped sensor exhibited a lower response upon aging, indicating deterioration over time (response$_{aged}$ < response$_{as\ prepared}$), (Figure S28a, Supporting Information). Conversely, the 50% v/v Mn-doped system remained quite stable with nearly zero percentage response difference (Figure 5b, blue points and S28b, Supporting Information), while the 80% Mn-doped sensor showed a significant improvement in its response with a positive response difference (Figure 5b magenta points, Figure S28c, Supporting Information). These results suggest a correlation between the response stability over time and the halide content and doping level.

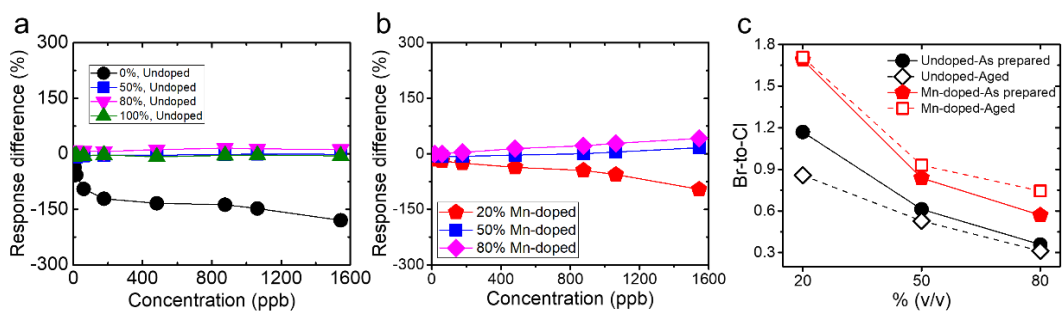

**Figure 5.** Percentage response difference between the as prepared and aged a) undoped and b) Mn-doped perovskite-based sensors as a function of the $O_3$ concentration. c) Br-to-Cl ratio of as prepared (filled) and aged (hollow) undoped (circles) and Mn-doped (pentagons) perovskites.

It is noteworthy that, compared to $CsPbBr_3$-based sensor which exhibited a significant response decrease upon time with a negative percentage difference of 150% (Figure 5a, black points and S26a, Supporting Information), both undoped and Mn-doped- based mixed halide sensors appeared more stable in terms of $O_3$ response over time. Moreover, although the initial observation might suggest an overall advantage in terms of highest response and stability of the pure $CsPbCl_3$-based sensor (Figure 5a, S26b, Supporting Information) compared to the mixed halide-based sensors (Figure S27 and 28, Supporting Information), a closer examination revealed its inability to distinguish different ozone concentrations, as evidenced by the similarity in oxidation curves (Figure S23b, Supporting Information). Notably, the 50% v/v Mn-doped sensor emerged as the sensor with the optimal combination of the highest response and stability among all the sensors tested (Figure 5b and S28b, Supporting Information). Despite this sensor having similar a Br-to-Cl ratio with the 20% v/v undoped sensor, as previously discussed (Figure 2b), this comparison provides clear evidence that Mn-doping improves response and sensing stability over time.

To understand the variations in the sensing behavior of both undoped and Mn-doped perovskite-based sensors over time (as-prepared and aged), a series of repetitive structural and morphological characterizations were carried out. Starting from the undoped perovskite sensors, SEM imaging revealed a well-preserved surface morphology of all the μCs over time (Figure S29, Supporting Information), while the XRD patterns and the absorbance spectra (Figure S30-S31, Supporting Information) revealed no substantial difference between the as-prepared and aged samples. However, a decrease in the Br-to-Cl ratio, determined by XPS survey spectra, was observed in all the aged samples compared to the as-prepared ones (Figure 5c and S32, Supporting Information), indicating the formation of Br-vacancies on the surface of the materials over time. The most significant difference in Br-to-Cl ratio between the as-prepared and aged sensor was observed for the 20% undoped sensor (Figure 5c), confirming that this system exhibits the lowest stability over time (Figure S27a, Supporting Information).

Similarly, $CsPbBr_3$ μCs showed a reduction in the Br-to-Pb ratio by 4.3% compared to the initial composition, indicating Br deficiency, related to the formation of Br vacancies on the surface of the material over time (Figure S33a, Supporting Information). It is worth noting that halide vacancies are the most mobile charges in halide perovskites due to their low migration activation energies, and are also more likely to move towards the surface of the perovskite material upon time.[48,49] The formation of these vacancies may induce ion migration, a process that potentially contributes to the observed increase in conductivity when voltage is applied, explaining the increased current intensity after exposing the perovskites to ambient conditions.[50,51,52] On the other hand, $CsPbCl_3$ μCs showed a 4.5% increase in the Cl content on the surface of the material over time (Figure S33b, Supporting Information). This redistribution of Cl content on the surface of the μCs could explain their enhanced conductivity, while the absence of extra vacancies on the surface could explain the stability in response of this sensor upon aging.

To further assess the fine structure of specific peaks of interest of the undoped systems, XPS narrow scan were conducted. In particular, Cs 3d, Pb 4f, and Br 3d chemical states were investigated (Figure S34-37, Supporting Information). After exposing the samples to ambient conditions for one month, no change in Cs 3d peaks were detected in all undoped perovskite systems indicating the low bonding interactions of Cs with Br and Cl ions either on surface or crystal lattice (Figure S34a, Supporting Information).[53] No change was observed for both $CsPbBr_3$ and $CsPbCl_3$ μCs as well (Figure S34b and c, Supporting Information).

Furthermore, the $f_{7/2}$ and $f_{5/2}$ chemical states of Pb 4f were deconvoluted into doublets as depicted in Figure S35 (Supporting Information). The peaks at approximately 138.1 eV and 143 eV correspond to the $Pb^{2+}$ species originating from Pb-X bonds, while no traces of metallic $Pb^0$ were observed, even after the prolonged exposure of the μCs to ambient conditions (Figure S35 a, Supporting Information). Additionally, a red-shift of Pb 4f doublets in the binding energy of 20% v/v undoped perovskite μCs was observed, implying modifications in the local environment of $[PbX_6]^{4-}$ octahedral (Figure S35c, Supporting Information). These changes may arise from the length of Pb-X bonds, which are associated with the strength of Pb-X interactions, possibly originating from halide migration or the formation of surface defects.[38,54, 55] On the contrary, the binding energy in Pb 4f of 50% and 80% v/v undoped μCs and $CsPbCl_3$ did not display any significant variations over time (Figure S35 b and c, Supporting Information), indicating the high stability of Cl-rich species towards ambient air and light, a finding in accordance with previous predictions by Cai et al. [56] Similar trends were observed in Br 3d and Cl 2p spectra (Figure S36-37, Supporting Information). In particular, it was noticed that the undoped systems exhibited a decrease in the relative intensity between Br 3d and Cs 4d peaks, suggesting further the reduction in Br content on the surface of each sensor and possibly the formation of Br vacancies (Figure S36, Supporting Information). It is noteworthy that among all the undoped systems the 20% v/v exhibited a more pronounced decrease compared to the other two, while the 50% v/v showed a more substantial decrease than the 80% v/v, indicating a concentration-dependent effect (Figure 5c). This observation is in consistent with the decreased Br-to-Cl ratio over time and the change in the sensing behavior of 20% and 50% v/v sensors towards p-type conductivity over time.

In the case of Mn-doped systems, SEM imaging revealed that the μCs retain their morphological features (Figure S38, Supporting Information), however significant differences were observed both structurally and optically. Particularly, the XRD analysis revealed an increase in the intensity of the secondary non-perovskite $CsPb_2X_5$ byproducts over time, which was increased by increasing the volume-to-volume ratio (Figure S39, Supporting Information). Apart from the formation of impurities after the prolonged exposure of the Mn-doped sensors to ambient conditions, a careful examination of the XRD patterns revealed peak splitting at ~15.5º, 21.8º, 31.3º and 35º (Figure 6a), that indicated symmetry-breaking and subsequently the phase transformation from their initial sub-tetragonal phase into the more energetically favorable orthorhombic crystal structure due to the incorporation of Cl anions.

Interestingly, the structural properties of both Mn-doped systems were re-evaluated after one year revealing more pronounced peak splitting for all samples with different v/v ratios (Figure S39, Supporting Information). Previous studies suggested that this transition resulted from the relatively soft crystal lattice of these materials, which caused intrinsic disorder phenomena, such as vacancy and dislocation formation.[57,58] These structural imperfections render halide perovskites highly sensitive to external stimuli such as light, water and oxygen.[59,60] Consequently, this susceptibility accelerated the formation and migration of ion defects, resulting to the distortion of crystal lattice by impacting the tilting of $[PbX_6]^{4-}$ octahedra and the subsequent reduction of its symmetry.[61,62] This finding corroborated further the interaction of $CsPbX_3$ μCs with atmospheric species and therefore the formation of $CsPb_2X_5$ impurities (Figure 7a). Additionally, regarding the Mn-doped systems, the introduction of dopants accelerated the deterioration process, possibly due the large radius mismatch between $Pb^{2+}$ and $Mn^{2+}$.[63]

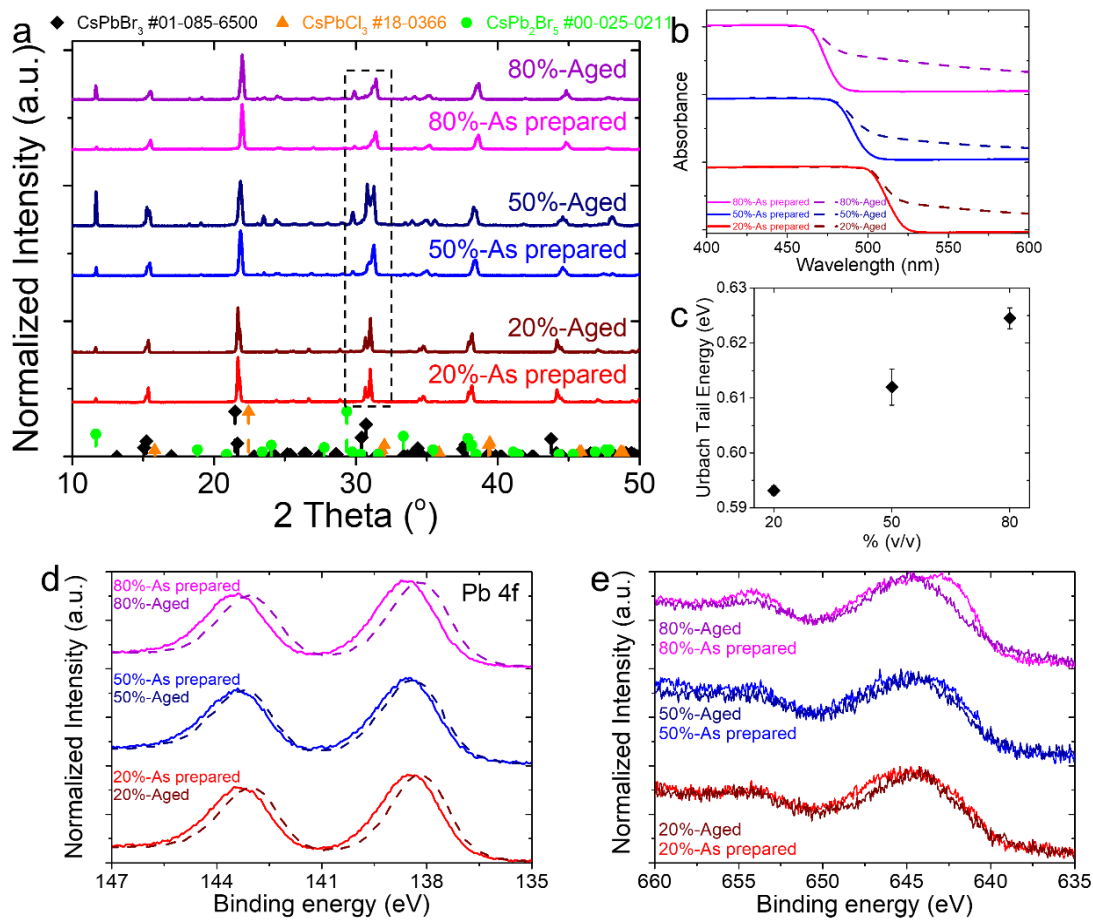

**Figure 6.** a) XRD patterns of the, as prepared and aged after their exposure to ambient conditions for a month, Mn-doped mixed halide perovskite μCs by varying the v/v ratio. b) Absorbance spectra of the same as-prepared (solid lines) and aged samples (dashed lines). c) Calculated Urbach energy as a function of volume-to-volume ratio. d-e) High resolution XPS spectra of Pb 4f and Mn 2p respectively.

Further insights into the effects of the prolonged exposure to ambient conditions of the Mn-doped systems were obtained through UV-Vis spectroscopy (Figure 6b). The absorbance spectra of the Mn-doped systems revealed an exponential decrease with decreasing energy over time, which can be described by the Urbach rule equation[64], while the calculation of Urbach energy showed an increase with higher doping level (Figure 6c). The presence of this Urbach tail could be linked to the presence of localized states in the bandgap, caused by imperfections within the crystal lattice related to dynamic disorder.[65] This dynamic disorder makes halide perovskites susceptible to intrinsic defects, inducing structural instabilities that were further confirmed previously by XRD (Figure 6a). These instabilities could possibly arise from the redistribution of Mn dopants over time, leading to changes in the electronic structure and absorbance of the materials, or a possible degradation resulting from exposure to various environmental factors, including photoxidation, known to influence the stability of Mn-doped nCs.[32]

The binding energies of Cs, Pb, Br and Cl were also examined to determine surface alterations that have occurred in each Mn-doped perovskite material over time. Similar to the undoped systems, Mn-doped mixed halide perovskite μCs did not exhibit any chemical changes in Cs 3d doublets, however a red-shift of Pb 4f and Cl 2p doublets in the binding energy was observed for all Mn-doped systems (Figure 6d and S40-41, Supporting Information). Moreover, even though no changes in the relative intensities between Br 3d and Cs 4d peaks were observed over time, an increasing Br-to-Cl ratio was found with increasing the Mn-content, indicating the possible formation of Cl vacancies (Figure 6c and S42-43, Supporting Information). It is of particular interest the remarkable increase in the Br-to-Cl ratio observed in the 80% v/v Mn-doped system, compared to the 20% and 50% counterparts, aligning with the significant increase in response over time (Figure 6c). Additionally, Mn 2p narrow scans revealed a decrease of $Mn^{2+}$ on the surface of Mn-doped systems after their prolonged exposure to ambient conditions, as shown in Figure 6e. This observation, along with the formation of Cl vacancies on the surface of the μCs, could possibly suggest that $MnCl_2$ may have migrated deeper into the crystal lattice through a defect-mediated diffusion that could affect the stability and ionic conductivity. It is noteworthy that, despite the observed instabilities of Mn-doped mixed halide perovskites over time, the capability of the Mn-doped perovskite- sensors to detect $O_3$ was remained. Additionally, despite the deterioration, the 50% v/v Mn-doped based-sensor demonstrated the best response stability over time.

## 3. Conclusion

This study highlights the critical role of metal cations and halide anions in $O_3$ detection, revealing that ligand-free undoped, and Mn-doped mixed halide $CsPbBr_{3-x}Cl_x$ perovskites exhibit varying responses based on halide content, with bromine-rich sensors showing p-type behavior, while chlorine-rich ones exhibit n-type behavior. Surface defects, particularly bromine and chlorine vacancies, are vital for oxygen adsorption and ozone sensing, as revealed by DFT calculations. In addition, Mn-doping enhances sensor performance by facilitating the $O_3$ absorption process. Intriguingly, deviations in sensing behavior are observed in Mn-doped systems compared to the undoped based sensors, suggesting a potential overlap in the roles of Mn and Br ions over Cl ions. Over time, Mn-doped sensors demonstrate improved stability and sensing capabilities, while the response of pure $CsPbBr_3$ sensors declines. Undoped mixed halides and $CsPbCl_3$ sensors maintain a stable but low response, with limited differentiation between $O_3$ levels. In contrast, Mn-doped sensors preserve their sensing abilities, with the 50% v/v Mn-doped sensor proving to be the best in terms of response and stability. The room-temperature-operating sensors fabricated from Mn-doped perovskites exhibited a reduced response in comparison to the advanced metal oxide sensors that require light irradiation.[66–68] However, their ability to operate effectively and recover at room temperature represents a significant advantage. Furthermore, when assessing stability, the manganese-doped perovskite sensors demonstrate superior stability relative to the previously reported organic-inorganic metal halide perovskite-based sensors.[27]

These findings offer valuable insights into the gas detection capabilities of halide perovskites, showcasing the intricate relationship between metal ions, halides, and surface defects. By manipulating halide composition and manganese doping levels, sensor response and stability can be tailored, paving the way for affordable, sensitive, and energy-efficient gas sensors with potential for selective detection. Moreover, the extended analysis of these sensors not only advances gas detection technologies but also provides vital information for enhancing the long-term stability and reliability of perovskite materials across various applications, including photovoltaics and optoelectronics. The demonstration of tunable sensitivity in mixed halide perovskites, coupled with the enhanced performance achieved through doping, opens up new avenues for designing bespoke sensing materials. The ability to control bandgap and defect concentrations through compositional engineering enables the optimization of sensor response to specific ozone concentrations or other target gases. Moreover, the insights gained into degradation pathways and long-term stability are paramount for developing durable and reliable sensors for real-world deployment.

## 4. Experimental Section/Methods

*Chemicals*: The chemical precursors CsBr (99.999% trace metals basis), $PbBr_2$ (≥98% trace metals basis), CsCl (≥99.5% trace metals basis), $PbCl_2$ (99.999% trace metals basis), and $MnCl_2$ (≥99% trace metals basis) were purchased from Sigma Aldrich and used without further treatment before the synthesis of the metal halide perovskite μCs. The solvents N, N-Dimethylformamide (DMF, ≥99.8%), toluene (≥99.5%) and Dimethyl Sulfoxide (DMSO, ≥99.7%) were purchased from Honeywell, Sigma-Aldrich and the Merck, respectively.

*Synthesis of $CsPbBr_3$ μCs*: Ligand-free $CsPbBr_3$ μCs were synthesized via a room temperature re-precipitation-based approach. A precursor solution was prepared by dissolving 0.4 mmol of CsBr and 0.4 mmol of $PbBr_2$ in 10 ml of DMF under ambient conditions. Subsequently, 1 ml of this precursor solution was added into 2 ml of toluene and the color of the solution turned instantly into bright yellow, indicating the rapid nucleation and crystal formation. Following this, the mixture was sonicated for 30 minutes to achieve a uniform μCs' size distribution.

*Synthesis of CsPbCl$_3$ μCs*: Similar to the CsPbBr$_3$ μCs, CsPbCl$_3$ μCs were formed by adding 1ml of the stock solution into 2 ml of toluene, followed by a 30-minute sonication. In this case, the precursor solution was prepared by dissolving 0.4 mmol of CsCl and 0.4 mmol of PbCl$_2$ in 10 ml of DMSO.

*Synthesis of undoped CsPbBr$_{3-x}$Cl$_x$ μCs*: CsPbBr$_{3-x}$Cl$_x$ μCs were synthesized through a typical anion exchange reaction. Initially, a PbCl$_2$ precursor solution was prepared by dissolving 0.6 mmol of PbCl$_2$ in 5 ml DMF. Following this, four distinct mixtures were prepared by adding precise amounts of PbCl$_2$ into the pre-synthesized CsPbBr$_3$ μCs with volume-to-volume ratio of 10%, 20%, 50% and 80% v/v. To accelerate the anion exchange process, the μCs were placed in an ultrasonic bath for 30 minutes.

*Synthesis of Mn-doped CsPbBr$_{3-x}$Cl$_x$ μCs*: Mn-doped CsPbBr$_{3-x}$Cl$_x$ μCs were synthesized via a modified post-synthetic halide exchange-driven cation exchange strategy (HEDCE) at room temperature.[37] A stock solution of MnCl$_2$ precursor was synthesized by dissolving 0.6 mmol of MnCl$_2$ powder into 5 ml of DMF. Subsequently, precise volumes of the stock MnCl$_2$ solution were added into the pre-formed of CsPbBr$_3$ μCs and the mixture was sonicated for 30 minutes. The doping level was controlled by using the same volume-to-volume ratios as those employed for the undoped mixed halides μCs, which was 20%, 50% and 80% v/v.

*Characterization of the materials*: Scanning Electron Microscopy (SEM, JEOL 7000) incorporated with Energy Dispersive X-Ray Spectrometer (EDS, INCA PentaFET-x3) operating at 20 kV was employed to analyze the surface morphology and elemental composition of the fabricated microcrystals. The crystal structures were investigated by X-Ray Diffraction (XRD, Bruker AXS D8 Advance copper anode diffractometer) over the 2θ collection range of 10º to 50º with scan rate of 0.02º/s, using a monochromatic Cu Kα radiation source (λ=1.54056 Å). SEM and XRD samples were prepared by drop-casting the μCs on indium tin oxide (ITO) coated glass substrates and glass substrates, respectively. The relative concentration Mn/Pb in Mn-doped mixed halide systems was estimated by Inductively Coupled Plasma Mass Spectroscopy (ICP-MS, Perkin Elmer NexION 350D). The microcrystals were in powder form and were diluted in 3 mL of 2% HNO$_3$ and heated at 95 °C overnight on heating blocks. The absorbance spectra of all the fabricated materials were measured by UV-Visible spectroscopy (Perkin Elmer Lambda 950 UV/Vis/NIR) over the range of 360 nm to 650 nm after drop-casting the microcrystals on glass substrates. The chemical compositions and electronic states of the perovskite-based microcrystals, drop-casted on ITO coated glass substrates, were determined by X-Ray Photoelectron Spectroscopy (XPS, SPECS-Germany, FlexMod) equipped with a PHOIBOS 100 1D-DLD energy analyzer and an Al Kα monochromatic X-Ray source (1486.7 eV) operated with 200 W and 15 kV. The recorded survey spectra were the following: C 1s, Pb 4f, Br 3d, Cs 3d, Cl 2p, Mn 2p and O 1s which recorded at a pass energy of 30 eV. All the binding energies in XPS spectra were calibrated using the C 1s at 284.8 eV as a reference. The data analysis was performed with SpecsLab Prodigy and CasaXPS (Casa software Ltd, DEMO version). A Tougaard baseline was used in combination with a Gaussian-Lorentzian function for the high-resolution spectra deconvolution, which was executed by OriginPro 2016 software.

*Preparation of the electrodes and gas sensing experiments*: The perovskite μCs were drop-casted onto commercially available interdigitated platinum electrodes on glass substrate (IDEs, 5 microns lines and gaps/glass substrate, Metrohm) and were left to dry under vacuum.

The gas sensing measurements were conducted at room temperature, in a custom-made gas sensing chamber, providing a controlled environment for the electrical assessment of the sensors. The gas sensing setup consists of certified gas suppliers connected with mass flow controllers, coupled with a stainless chamber which was initially evacuated down to $10^{-3}$ mbar. To evaluate the ozone sensing capability of all fabricated materials, a commercial ozone analyzer (Thermo Electron Corporation, Model 49i) was employed. The analyzer was used to supply and record accurately controlled ozone concentrations that were introduced in the chamber at a flow rate of 500 sccm (standard cm$^3$/min) flow. Electrical current measurements over time were executed using an electrometer (Keithley 6517A) by applying a constant voltage. The values of the voltage ranged from 1 to 3.5 V, depending on the conductivity of each sensor. The sensing process was initiated by exposing the sensors to ozone gas for 150 s, followed by a 200 s recovery with synthetic air. All samples were exposed to different ozone concentrations ranging from 1567 down to 4 ppb. During the experimental procedure, the pressure in the chamber maintained constant to 700 mbar.

The performance of the sensors for each gas is assessed through their response (S), defined as:

$S = \frac{I_{gas}}{I_{air}}$ or $S = \frac{I_{air}}{I_{gas}}$   (Equation 1)

where I$_{gas}$ and I$_{air}$ are the electrical current values after saturation in the presence of reducing/oxidizing gas and synthetic air, respectively.

The percentage response difference between the as prepared and aged MCs' samples were calculated by the formula:

Percentage response difference = $\frac{r_{aging} - r_{as-prepared}}{\left[\frac{r_{aging} + r_{as-prepared}}{2}\right]} \times 100$ (Equation 2)

*Computational methodology*: The (001) oriented CsPbBr$_3$ surfaces were simulated by five-layer slab models, as suggested by W. Xing et.al., built with a 2×2 supercell of the bulk orthorhombic phase of CsPbBr$_3$ with atom positions and lattice parameters taken from the experimental data of C.C. Stoumpos et al..[42,69] From this bulk orthorhombic structure, symmetric slabs consisting of alternating PbBr$_2$ and CsBr layers were built, terminated by PbBr$_2$ planes. The constructed slabs with lattice constants a=11.6264 Å, b=11.7351 Å, c=32.6424 Å are finite in the z-direction, separated by a vacuum gap of 20 Å, and periodic in x- and y- directions. Modified surfaces with Br and Pb vacancy were constructed by removing 1 Br and 1 Pb atom from the PbBr$_2$ termination plane, respectively, while Mn-doped CsPbBr$_3$ surface was constructed by substituting 1 Pb atom with 1 Mn atom, and Cl-doped by substituting 1 Br atom with 1 Cl atom, or all Br atoms from the uppermost layer of the simulated surface with Cl atoms. The (001) oriented CsPbCl$_3$ surface presented in the SI section was constructed by using the bulk tetragonal structure with lattice constants obtained from the Materials Project database[70], while the corresponding defected surface with a Cl vacancy was constructed by removing 1 Cl atom from the PbCl$_2$ termination plane.

DFT calculations were performed as implemented in the Vienna ab initio simulation package (VASP).[71,72] The generalized gradient approximation (GGA) with the Perdew-Becke-Erzenhof (PBE) parametrization for the exchange-correlation (XC) was employed, while interactions of valence electrons with the remaining ions were modeled within the projector augmented wave (PAW) formalism.[73,74] An energy cut-off of 400 eV for the plane wave expansion was used and van der Waals interactions were taken into account employing the DFT-D3 scheme of Grimme.[75] An energy convergence criterion of 10$^{-6}$ eV was used for the relaxation of the electronic degrees of freedom and all structures were considered fully relaxed when the maximum force acting on each atom was less than 0.01 eV·Å$^{-1}$. For the Brillouin zone integrations, a 4×4×1 Γ-centered k-point mesh was selected and a Gaussian smearing of 0.1 eV was used. During structure optimization, the cell volume and shape were not allowed to change, the positions of the atoms in the bottom three layers of the slabs were held fixed, while the atoms in the top two layers and of the adsorbate were allowed to relax.[76]

The adsorption energies ΔE$_{ads}$ were calculated as $\Delta E_{ads} = E_{O2+slab} - E_{slab} - E_{O2}$, where $E_{O2+slab}$ is the energy of the slab with the adsorbed O$_2$ molecule, $E_{slab}$ is the energy of the slab, and $E_{O2}$ is the spin-polarized energy of the O$_2$ molecule in vacuum.

Charge density difference plots were calculated as the difference of the electron density of the slab + O$_2$ complex system minus the sum of the isolated slab and O$_2$ within the conformation of the complex, and were plotted using VESTA software on a 0.005 (e/bohr$^3$) isosurface.[77]

**Supporting Information**

Supporting Information is available from the Wiley Online Library or from the author.

**Acknowledgements**

This research project has received funding from the EU's Horizon Europe framework programme for research and innovation under grant agreement BRIDGE (n. 101079421 from 01/10/2022 – 30/9/2025). In addition, FLAG-ERA Joint Transnational Call 2019 for transnational research projects in synergy with the two FET Flagships Graphene Flagship & Human Brain Project - ERA-NETS 2019b (PeroGaS: MIS 5070514) is acknowledged for the financial support. We would like also to thank Mrs Alexandra Manousaki for the observation of the samples on SEM, Dr Emmanouel Spanakis for conducting the XPS measurements, the Electron Microscopy Laboratory of the University of Crete for providing access to HRTEM and SEM facilities and the Material Science and Technology Department of the University of Crete for providing access to XPS facilities. This work was also supported by NFFA EUROPE Pilot (EU H2020 framework programme) under grant agreement no.101007417 from 1/03/2021 to 28/02/2026 for the NFFA 275 ID proposal at the Joint Research Center-Ispra with Dr Pascal Colpo, Dr Ivana Bianchi and Dr Otmar Geiss responsible for the ICP-MS measurements. The research work regarding the DFT calculations was supported by the Hellenic Foundation for Research and Innovation (HFRI) under the 3rd Call for HFRI PhD Fellowships (Fellowship Number: 5706). R.M. Giappa and I. Remediakis acknowledge computational time granted from the National Infrastructures for Research and Technology S.A. (GRNET S.A.) in the National HPC facility, ARIS, under projects NANOPTOCAT and CompNanoMat.

**Conflict of Interest**

All authors must declare financial/commercial conflicts of interest.

**References**

(1) Manisalidis, I.; Stavropoulou, E.; Stavropoulos, A.; Bezirtzoglou, E. Environmental and Health Impacts of Air Pollution: A Review. *Front. Public Heal.* **2020**, *8* (February), 1–13. https://doi.org/10.3389/fpubh.2020.00014.

(2) Pénard-Morand, C.; Annesi-Maesano, I. Air Pollution: From Sources of Emissions to Health Effects. *Breathe* **2004**, *1* (2), 108–119. https://doi.org/10.1183/18106838.0102.108.

(3) Seiyama, T.; Kato, A.; Fujiishi, K.; Nagatani, M. A New Detector for Gaseous Components Using Semiconductive Thin Films Determination of Benzoic Acid in Phthalic Anhydride by Gas Liquid Chromatography. *Anal. Chem.* **1962**, *34* (11), 1502–1503.

(4) Korotcenkov, G. The Role of Morphology and Crystallographic Structure of Metal Oxides in Response of Conductometric-Type Gas Sensors. *Mater. Sci. Eng. R Reports* **2008**, *61* (1–6), 1–39. https://doi.org/10.1016/j.mser.2008.02.001.

(5) Zhou, T.; Zhang, T. Recent Progress of Nanostructured Sensing Materials from 0D to 3D: Overview of Structure–Property-Application Relationship for Gas Sensors. *Small Methods* **2021**, *5* (9), 1–32. https://doi.org/10.1002/smtd.202100515.

(6) Govardhan, K.; Nirmala Grace, A. Metal/Metal Oxide Doped Semiconductor Based Metal Oxide Gas Sensors - A Review. *Sens. Lett.* **2016**, *14* (8), 741–750. https://doi.org/10.1166/sl.2016.3710.

(7) Chai, H.; Zheng, Z.; Liu, K.; Xu, J.; Wu, K.; Luo, Y.; Liao, H.; Debliquy, M.; Zhang, C. Stability of Metal Oxide Semiconductor Gas Sensors: A Review. *IEEE Sens. J.* **2022**, *22* (6), 5470–5481. https://doi.org/10.1109/JSEN.2022.3148264.

(8) Korotcenkov, G.; Cho, B. K. Engineering Approaches to Improvement of Conductometric Gas Sensor Parameters. Part 2: Decrease of Dissipated (Consumable) Power and Improvement Stability and Reliability. *Sensors Actuators, B Chem.* **2014**, *198*, 316–341. https://doi.org/10.1016/j.snb.2014.03.069.

(9) Moseley, P. T. Progress in the Development of Semiconducting Metal Oxide Gas Sensors: A Review. *Meas. Sci. Technol.* **2017**, *28* (8). https://doi.org/10.1088/1361-6501/aa7443.

(10) Li, Z.; Li, H.; Wu, Z.; Wang, M.; Luo, J.; Torun, H.; Hu, P.; Yang, C.; Grundmann, M.; Liu, X.; Fu, Y. Advances in Designs and Mechanisms of Semiconducting Metal Oxide Nanostructures for High-Precision Gas Sensors Operated at Room Temperature. *Mater. Horizons* **2019**, *6* (3), 470–506. https://doi.org/10.1039/c8mh01365a.

(11) Huang, Y.; Zhang, J.; Zhang, X.; Jian, J.; Zou, J.; Jin, Q.; Zhang, X. The Ammonia Detection of Cesium Lead Halide Perovskite Quantum Dots in Different Halogen Ratios at Room Temperature. *Opt. Mater. (Amst).* **2022**, *134* (PA), 113155. https://doi.org/10.1016/j.optmat.2022.113155.

(12) Ayesh, A. I.; Alghamdi, S. A.; Salah, B.; Bennett, S. H.; Crean, C.; Sellin, P. J. Results in Physics High Sensitivity H 2 S Gas Sensors Using Lead Halide Perovskite Nanoparticles. *Results Phys.* **2022**, *35* (December 2021), 105333.

https://doi.org/10.1016/j.rinp.2022.105333.

(13) Parfenov, A. A.; Yamilova, O. R.; Gutsev, L. G.; Sagdullina, D. K.; Novikov, A. V; Ramachandran, B. R.; Stevenson, K. J.; Aldoshin, S. M.; Troshin, P. A. Highly Sensitive and Selective Ammonia Gas Sensor Based on FAPbCl 3 Lead Halide Perovskites. *J. Mater. Chem. C* **2021**, *9* (7), 2561–2568. https://doi.org/10.1039/D0TC03559A.

(14) Casanova-chafer, J.; Llobet, E. ChemComm The Role of Anions and Cations in the Gas Sensing Mechanisms of Graphene Decorated with Lead Halide Perovskite Nanocrystals †. **2020**, 8956–8959. https://doi.org/10.1039/d0cc02984j.

(15) Gagaoudakis, E.; Panagiotopoulos, A.; Maksudov, T.; Moschogiannaki, M.; Katerinopoulou, D.; Kakavelakis, G.; Kiriakidis, G.; Binas, V.; Kymakis, E.; Petridis, K. Self-Powered, Flexible and Room Temperature Operated Solution Processed Hybrid Metal Halide p-Type Sensing Element for Efficient Hydrogen Detection. *JPhys Mater.* **2020**, *3* (1). https://doi.org/10.1088/2515-7639/ab60c3.

(16) Liu, X.; Cheng, S.; Liu, H.; Hu, S.; Zhang, D.; Ning, H. A Survey on Gas Sensing Technology. *Sensors (Switzerland)* **2012**, *12* (7), 9635–9665. https://doi.org/10.3390/s120709635.

(17) Fu, X.; Jiao, S.; Dong, N.; Lian, G.; Zhao, T.; Lv, S.; Wang, Q.; Cui, D. A CH3NH3PbI3 Film for a Room-Temperature NO2 Gas Sensor with Quick Response and High Selectivity. *RSC Adv.* **2018**, *8* (1), 390–395. https://doi.org/10.1039/c7ra11149e.

(18) Huang, L.; Ge, Z.; Zhang, X.; Zhu, Y. Oxygen-Induced Defect-Healing and Photo-Brightening of Halide Perovskite Semiconductors: Science and Application. *J. Mater. Chem. A* **2021**, *9* (8), 4379–4414. https://doi.org/10.1039/d0ta10946k.

(19) Stoeckel, M. A.; Gobbi, M.; Bonacchi, S.; Liscio, F.; Ferlauto, L.; Orgiu, E.; Samorì, P. Reversible, Fast, and Wide-Range Oxygen Sensor Based on Nanostructured Organometal Halide Perovskite. *Adv. Mater.* **2017**, *29* (38), 1–7. https://doi.org/10.1002/adma.201702469.

(20) Shellaiah, M.; Sun, K. W. Review on Sensing Applications of Perovskite Nanomaterials. *Chemosensors* **2020**, *8* (3), 55. https://doi.org/10.3390/chemosensors8030055.

(21) Viebahn-hänsler, R.; Sonia, O.; Fernández, L.; Fahmy, Z.; Sonia, O.; Fernández, L.; Fahmy, Z.; Viebahn-hänsler, R.; Sonia, O.; Fernández, L.; Fahmy, Z. Ozone : Science & Engineering The Journal of the International Ozone Association Ozone in Medicine :

The Low-Dose Ozone Concept — Guidelines and Treatment Strategies Ozone in Medicine : The Low-Dose Ozone Concept — Guidelines and Treatment Strategies. **2012**, *9512*. https://doi.org/10.1080/01919512.2012.717847.

(22) Vozmilov, A. G.; Yu, I. R. The Usage of Ozone in Agriculture Technological Processes. **2016**, 0–3.

(23) Filippidou EC; Koukouliata A. Ozone Effects on the Respiratory System. *Prog Heal. Sci* **2011**, *1* (2), 144–155.

(24) Karmakar, S. P.; Das, A. B.; Gurung, C.; Ghosh, C. Effects of Ozone on Plant Health and Environment: A Mini Review. *Res. J. Agric. Sci.* **2022**, *13* (3), 612–619.

(25) Update, G. Air Quality Guidelines. **2005**.

(26) Iqbal, M. M.; Muhammad, G.; Hussain, M. A.; Hanif, H.; Raza, M. A.; Shafiq, Z. Recent Trends in Ozone Sensing Technology. *Anal. Methods* **2023**, 2798–2822. https://doi.org/10.1039/d3ay00334e.

(27) Kakavelakis, G.; Gagaoudakis, E.; Petridis, K.; Petromichelaki, V.; Binas, V.; Kiriakidis, G.; Kymakis, E. Solution Processed CH3NH3PbI3-XClx Perovskite Based Self-Powered Ozone Sensing Element Operated at Room Temperature. *ACS Sensors* **2018**, *3* (1), 135–142. https://doi.org/10.1021/acssensors.7b00761.

(28) Argyrou, A.; Brintakis, K.; Kostopoulou, A.; Gagaoudakis, E.; Demeridou, I.; Binas, V.; Kiriakidis, G.; Stratakis, E. Highly Sensitive Ozone and Hydrogen Sensors Based on Perovskite Microcrystals Directly Grown on Electrodes. *J. Mater.* **2021**, No. xxxx. https://doi.org/10.1016/j.jmat.2021.07.002.

(29) Brintakis, K.; Gagaoudakis, E.; Kostopoulou, A.; Faka, V.; Argyrou, A.; Binas, V.; Kiriakidis, G.; Stratakis, E. Ligand-Free All-Inorganic Metal Halide Nanocubes for Fast, Ultra-Sensitive and Self-Powered Ozone Sensors. *Nanoscale Adv.* **2019**, *1* (7), 2699–2706. https://doi.org/10.1039/c9na00219g.

(30) Chen, X.; Han, D.; Su, Y.; Zeng, Q.; Liu, L.; Shen, D. Structural and Electronic Properties of Inorganic Mixed Halide Perovskites. *Phys. Status Solidi - Rapid Res. Lett.* **2018**, *12* (8), 1–6. https://doi.org/10.1002/pssr.201800193.

(31) Yan, L.; Wang, M.; Zhai, C.; Zhao, L.; Lin, S. Symmetry Breaking Induced Anisotropic Carrier Transport and Remarkable Thermoelectric Performance in Mixed Halide Perovskites CsPb(I1- XBrx)3. *ACS Appl. Mater. Interfaces* **2020**, *12* (36), 40453–40464. https://doi.org/10.1021/acsami.0c07501.

(32) Luo, B.; Li, F.; Xu, K.; Guo, Y.; Liu, Y.; Xia, Z.; Zhang, J. Z. B-Site Doped Lead Halide Perovskites: Synthesis, Band Engineering, Photophysics, and Light Emission

Applications. *J. Mater. Chem. C* **2019**, *7* (10), 2781–2808. https://doi.org/10.1039/c8tc05741a.

(33) Li, X.; Ma, J.; He, H. Recent Advances in Catalytic Decomposition of Ozone. *J. Environ. Sci. (China)* **2020**, *94*, 14–31. https://doi.org/10.1016/j.jes.2020.03.058.

(34) Hossain, K. M.; Hasan, M. Z.; Ali, M. L. Narrowing Bandgap and Enhanced Mechanical and Optoelectronic Properties of Perovskite Halides: Effects of Metal Doping. *AIP Adv.* **2021**, *11* (1). https://doi.org/10.1063/5.0039308.

(35) Lin, F.; Li, F.; Lai, Z.; Cai, Z.; Wang, Y.; Wolfbeis, O. S.; Chen, X. MnII-Doped Cesium Lead Chloride Perovskite Nanocrystals: Demonstration of Oxygen Sensing Capability Based on Luminescent Dopants and Host-Dopant Energy Transfer. *ACS Appl. Mater. Interfaces* **2018**, *10* (27), 23335–23343. https://doi.org/10.1021/acsami.8b06329.

(36) Bhatia, H.; Ghosh, B.; Debroye, E. Colloidal FAPbBr3 Perovskite Nanocrystals for Light Emission: What's Going On? *J. Mater. Chem. C* **2022**, *10* (37), 13437–13461. https://doi.org/10.1039/d2tc01373h.

(37) Huang, G.; Wang, C.; Xu, S.; Zong, S.; Lu, J.; Wang, Z.; Lu, C.; Cui, Y. Postsynthetic Doping of MnCl2 Molecules into Preformed CsPbBr3 Perovskite Nanocrystals via a Halide Exchange-Driven Cation Exchange. *Adv. Mater.* **2017**, *29* (29), 10–14. https://doi.org/10.1002/adma.201700095.

(38) Gualdrón-Reyes, A. F.; Rodríguez-Pereira, J.; Amado-González, E.; Rueda-P, J.; Ospina, R.; Masi, S.; Yoon, S. J.; Tirado, J.; Jaramillo, F.; Agouram, S.; Muñoz-Sanjosé, V.; Giménez, S.; Mora-Seró, I. Unravelling the Photocatalytic Behavior of All-Inorganic Mixed Halide Perovskites: The Role of Surface Chemical States. *ACS Appl. Mater. Interfaces* **2020**, *12* (1), 914–924. https://doi.org/10.1021/acsami.9b19374.

(39) Zhang, D.; Yang, Y.; Bekenstein, Y.; Yu, Y.; Gibson, N. A.; Wong, A. B.; Eaton, S. W.; Kornienko, N.; Kong, Q.; Lai, M.; Alivisatos, A. P.; Leone, S. R.; Yang, P. Synthesis of Composition Tunable and Highly Luminescent Cesium Lead Halide Nanowires through Anion-Exchange Reactions. *J. Am. Chem. Soc.* **2016**, *138* (23), 7236–7239. https://doi.org/10.1021/jacs.6b03134.

(40) Li, C.; Zhang, N.; Gao, P. Lessons Learned: How to Report XPS Data Incorrectly about Lead-Halide Perovskites. *Mater. Chem. Front.* **2023**, No. January. https://doi.org/10.1039/d3qm00574g.

(41) Liu, W.; Lin, Q.; Li, H.; Wu, K.; Robel, I.; Pietryga, J. M.; Klimov, V. I. Mn2+-Doped

Lead Halide Perovskite Nanocrystals with Dual-Color Emission Controlled by Halide Content. *J. Am. Chem. Soc.* **2016**, *138* (45), 14954–14961. https://doi.org/10.1021/jacs.6b08085.

(42) Xing, W.; Yao, Q.; Zhu, W.; Jiang, H.; Zhang, X.; Ji, Y.; Shao, J.; Xiong, W.; Wang, B.; Zhang, B.; Luo, X.; Zheng, Y. Donor–Acceptor Competition via Halide Vacancy Filling for Oxygen Detection of High Sensitivity and Stability by All-Inorganic Perovskite Films. *Small* **2021**, *2102733*, 2102733. https://doi.org/10.1002/smll.202102733.

(43) Chen, H.; Zhang, M.; Bo, R.; Barugkin, C.; Zheng, J.; Ma, Q.; Huang, S.; Ho-baillie, A. W. Y.; Catchpole, K. R.; Tricoli, A. Superior Self-Powered Room-Temperature Chemical Sensing with Light-Activated Inorganic Halides Perovskites. **2018**, *1702571*, 1–7. https://doi.org/10.1002/smll.201702571.

(44) Petromichelaki, E.; Gagaoudakis, E.; Moschovis, K.; Tsetseris, L.; Anthopoulos, T. D.; Kiriakidis, G.; Binas, V. Highly Sensitive and Room Temperature Detection of Ultra-Low Concentrations of O3 Using Self-Powered Sensing Elements of Cu2O Nanocubes. *Nanoscale Adv.* **2019**, *1* (5), 2009–2017. https://doi.org/10.1039/c9na00043g.

(45) Films, A. P. Donor – Acceptor Competition via Halide Vacancy Filling for Oxygen Detection of High Sensitivity and Stability by All-Inorganic Perovskite Films. **2021**. https://doi.org/10.1002/smll.202102733.

(46) Kabitakis, V.; Gagaoudakis, E.; Moschogiannaki, M.; Kiriakidis, G.; Seitkhan, A.; Firdaus, Y.; Faber, H.; Yengel, E.; Loganathan, K.; Deligeorgis, G.; Tsetseris, L.; Anthopoulos, T. D.; Binas, V. A Low-Power CuSCN Hydrogen Sensor Operating Reversibly at Room Temperature. *Adv. Funct. Mater.* **2022**, *32* (7), 1–9. https://doi.org/10.1002/adfm.202102635.

(47) Douloumis, A.; Vrithias, N. R. E.; Katsarakis, N.; Remediakis, I. N.; Kopidakis, G. Tuning the Workfunction of ZnO through Surface Doping with Mn from First-Principles Simulations. *Surf. Sci.* **2022**, *726* (June), 122175. https://doi.org/10.1016/j.susc.2022.122175.

(48) Chen, C.; Fu, Q.; Guo, P.; Chen, H.; Wang, M.; Luo, W.; Zheng, Z. Ionic Transport Characteristics of Large-Size CsPbBr3 Single Crystals. *Mater. Res. Express* **2019**, *6* (11). https://doi.org/10.1088/2053-1591/ab4d79.

(49) Yin, W. J.; Shi, T.; Yan, Y. Unusual Defect Physics in CH3NH3PbI3 Perovskite Solar Cell Absorber. *Appl. Phys. Lett.* **2014**, *104* (6). https://doi.org/10.1063/1.4864778.

(50) Senocrate, A.; Maier, J. Solid-State Ionics of Hybrid Halide Perovskites. *J. Am. Chem.*

*Soc.* **2019**, *141* (21), 8382–8396. https://doi.org/10.1021/jacs.8b13594.

(51) Walsh, A.; Stranks, S. D. Taking Control of Ion Transport in Halide Perovskite Solar Cells. *ACS Energy Lett.* **2018**, *3* (8), 1983–1990. https://doi.org/10.1021/acsenergylett.8b00764.

(52) Leupold, N.; Seibel, A. L.; Moos, R.; Panzer, F. Electrical Conductivity of Halide Perovskites Follows Expectations from Classical Defect Chemistry. *Eur. J. Inorg. Chem.* **2021**, *2021* (28), 2882–2889. https://doi.org/10.1002/ejic.202100381.

(53) Li, X.; Wu, Y.; Zhang, S.; Cai, B.; Gu, Y.; Song, J.; Zeng, H. CsPbX3 Quantum Dots for Lighting and Displays: Roomerature Synthesis, Photoluminescence Superiorities, Underlying Origins and White Light-Emitting Diodes. *Adv. Funct. Mater.* **2016**, *26* (15), 2435–2445. https://doi.org/10.1002/adfm.201600109.

(54) Huang, J.; Yan, H.; Zhou, D.; Zhang, J.; Deng, S.; Xu, P.; Chen, R.; Kwok, H. S.; Li, G. Introducing Ion Migration and Light-Induced Secondary Ion Redistribution for Phase-Stable and High-Efficiency Inorganic Perovskite Solar Cells. *ACS Appl. Mater. Interfaces* **2020**, *12* (36), 40364–40371. https://doi.org/10.1021/acsami.0c12068.

(55) Li, Z.; Xu, J.; Zhou, S.; Zhang, B.; Liu, X.; Dai, S.; Yao, J. CsBr-Induced Stable CsPbI 3- x Br x (x < 1) Perovskite Films at Low Temperature for Highly Efficient Planar Heterojunction Solar Cells. *ACS Appl. Mater. Interfaces* **2018**, *10* (44), 38183–38192. https://doi.org/10.1021/acsami.8b11474.

(56) Cai, M. Q.; Busipalli, D. L.; Xu, S. H.; Nachimuthu, S.; Jiang, J. C. Exploring the Air Stability of All-Inorganic Halide Perovskites in the Presence of Photogenerated Electrons by DFT and AIMD Studies. *Sustain. Energy Fuels* **2022**, *6* (16), 3778–3787. https://doi.org/10.1039/d2se00806h.

(57) Chen, J. K.; Zhao, Q.; Shirahata, N.; Yin, J.; Bakr, O. M.; Mohammed, O. F.; Sun, H. T. Shining Light on the Structure of Lead Halide Perovskite Nanocrystals. *ACS Mater. Lett.* **2021**, *3*, 845–861. https://doi.org/10.1021/acsmaterialslett.1c00197.

(58) Jin, H.; Debroye, E.; Keshavarz, M.; Scheblykin, I. G.; Roeffaers, M. B. J.; Hofkens, J.; Steele, J. A. It's a Trap! On the Nature of Localised States and Charge Trapping in Lead Halide Perovskites. *Mater. Horizons* **2020**, *7* (2), 397–410. https://doi.org/10.1039/c9mh00500e.

(59) Lou, S.; Xuan, T.; Wang, J. Stability: A Desiderated Problem for the Lead Halide Perovskites. *Opt. Mater. X* **2019**, *1* (April), 100023. https://doi.org/10.1016/j.omx.2019.100023.

(60) Ju, M. G.; Chen, M.; Zhou, Y.; Dai, J.; Ma, L.; Padture, N. P.; Zeng, X. C. Toward

Eco-Friendly and Stable Perovskite Materials for Photovoltaics. *Joule* **2018**, *2* (7), 1231–1241. https://doi.org/10.1016/j.joule.2018.04.026.

(61) Ma, J. P.; Yin, J.; Chen, Y. M.; Zhao, Q.; Zhou, Y.; Li, H.; Kuroiwa, Y.; Moriyoshi, C.; Li, Z. Y.; Bakr, O. M.; Mohammed, O. F.; Sun, H. T. Defect-Triggered Phase Transition in Cesium Lead Halide Perovskite Nanocrystals. *ACS Mater. Lett.* **2019**, *1* (1), 185–191. https://doi.org/10.1021/acsmaterialslett.9b00128.

(62) Ran, C.; Liu, X.; Gao, W.; Li, M.; Wu, Z.; Xia, Y.; Chen, Y. Healing Aged Metal Halide Perovskite toward Robust Optoelectronic Devices: Mechanisms, Strategies, and Perspectives. *Nano Energy* **2023**, *108* (December 2022), 108219. https://doi.org/10.1016/j.nanoen.2023.108219.

(63) Zhuang, B.; Liu, Y.; Yuan, S.; Huang, H.; Chen, J.; Chen, D. Glass Stabilized Ultra-Stable Dual-Emitting Mn-Doped Cesium Lead Halide Perovskite Quantum Dots for Cryogenic Temperature Sensing. *Nanoscale* **2019**, *11* (32), 15010–15016. https://doi.org/10.1039/c9nr05831a.

(64) Li, M.; Huang, P.; Zhong, H. Current Understanding of Band-Edge Properties of Halide Perovskites: Urbach Tail, Rashba Splitting, and Exciton Binding Energy. *J. Phys. Chem. Lett.* **2023**, *14* (6), 1592–1603. https://doi.org/10.1021/acs.jpclett.2c03525.

(65) Falsini, N.; Roini, G.; Ristori, A.; Calisi, N.; Biccari, F.; Vinattieri, A. Analysis of the Urbach Tail in Cesium Lead Halide Perovskites. *J. Appl. Phys.* **2022**, *131* (1). https://doi.org/10.1063/5.0076712.

(66) Joshi, N.; da Silva, L. F.; Shimizu, F. M.; Mastelaro, V. R.; M'Peko, J. C.; Lin, L.; Oliveira, O. N. UV-Assisted Chemiresistors Made with Gold-Modified ZnO Nanorods to Detect Ozone Gas at Room Temperature. *Microchim. Acta* **2019**, *186* (7). https://doi.org/10.1007/s00604-019-3532-4.

(67) Wu, C. H.; Chang, K. W.; Li, Y. N.; Deng, Z. Y.; Chen, K. L.; Jeng, C. C.; Wu, R. J.; Chen, J. H. Improving the Sensitive and Selective of Trace Amount Ozone Sensor on Indium-Gallium-Zinc Oxide Thin Film by Ultraviolet Irradiation. *Sensors Actuators, B Chem.* **2018**, *273* (March), 1713–1718. https://doi.org/10.1016/j.snb.2018.07.075.

(68) da Silva, L. F.; M'Peko, J. C.; Catto, A. C.; Bernardini, S.; Mastelaro, V. R.; Aguir, K.; Ribeiro, C.; Longo, E. UV-Enhanced Ozone Gas Sensing Response of ZnO-SnO2 Heterojunctions at Room Temperature. *Sensors Actuators, B Chem.* **2017**, *240*, 573–579. https://doi.org/10.1016/j.snb.2016.08.158.

(69) Stoumpos, C. C.; Malliakas, C. D.; Peters, J. A.; Liu, Z.; Sebastian, M.; Im, J.;

Chasapis, T. C.; Wibowo, A. C.; Chung, D. Y.; Freeman, A. J.; Wessels, B. W.; Kanatzidis, M. G. Crystal Growth of the Perovskite Semiconductor CsPbBr3: A New Material for High-Energy Radiation Detection. *Cryst. Growth Des.* **2013**, *13* (7), 2722–2727. https://doi.org/10.1021/cg400645t.

(70) Jain, A.; Ong, S. P.; Hautier, G.; Chen, W.; Richards, W. D.; Dacek, S.; Cholia, S.; Gunter, D.; Skinner, D.; Ceder, G.; Persson, K. A. Commentary: The Materials Project: A Materials Genome Approach to Accelerating Materials Innovation. *APL Mater.* **2013**, *1* (1). https://doi.org/10.1063/1.4812323.

(71) Kresse, G.; Furthmüller, J. Efficiency of Ab-Initio Total Energy Calculations for Metals and Semiconductors Using a Plane-Wave Basis Set. *Comput. Mater. Sci.* **1996**, *6* (1), 15–50. https://doi.org/10.1016/0927-0256(96)00008-0.

(72) Kresse, G.; Hafner, J. Ab Initio Molecular Dynamics for Liquid Metals. *Phys. Rev. B* **1993**, *47* (1), 558–561. https://doi.org/10.1103/PhysRevB.47.558.

(73) Perdew, J. P.; Burke, K.; Ernzerhof, M. Generalized Gradient Approximation Made Simple. *Phys. Rev. Lett.* **1996**, *77* (18), 3865–3868. https://doi.org/10.1103/PhysRevLett.77.3865.

(74) Joubert, D. From Ultrasoft Pseudopotentials to the Projector Augmented-Wave Method. *Phys. Rev. B - Condens. Matter Mater. Phys.* **1999**, *59* (3), 1758–1775. https://doi.org/10.1103/PhysRevB.59.1758.

(75) Grimme, S.; Antony, J.; Ehrlich, S.; Krieg, H. A Consistent and Accurate Ab Initio Parametrization of Density Functional Dispersion Correction (DFT-D) for the 94 Elements H-Pu. *J. Chem. Phys.* **2010**, *132* (15). https://doi.org/10.1063/1.3382344.

(76) Pack, J. D.; Monkhorst, H. J. "special Points for Brillouin-Zone Integrations"-a Reply. *Phys. Rev. B* **1977**, *16* (4), 1748–1749. https://doi.org/10.1103/PhysRevB.16.1748.

(77) Momma, K.; Izumi, F. VESTA: A Three-Dimensional Visualization System for Electronic and Structural Analysis. *J. Appl. Crystallogr.* **2008**, *41* (3), 653–658. https://doi.org/10.1107/S0021889808012016.

Table of content:

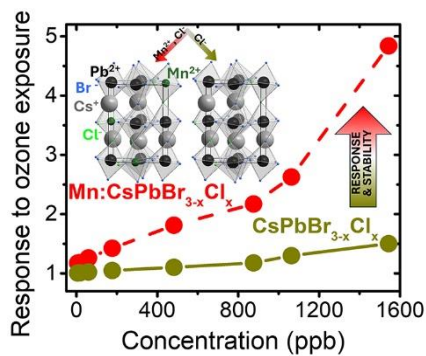

Mixed halide perovskites, especially Mn-doped, demonstrate a robust and sensitive response to $O_3$ at room temperature. Atomistic simulations and experiments confirm that Mn-doping boosts $O_3$ sensing by improving gas adsorption. The evaluation of sensing performance and long-term stability provides insights into the optimal halide combination and Mn-doping level, marking a breakthrough in the development of cost-effective, high-performance gas sensors.

# Supporting Information

**Towards the optimization of a perovskite-based room temperature ozone sensor: A multifaceted approach in pursuit of sensitivity, stability, and understanding of mechanism**

*Aikaterini Argyrou, Rafaela Maria Giappa, Emmanouil Gagaoudakis, Vasilios Binas, Ioannis Remediakis, Konstantinos Brintakis\*, Athanasia Kostopoulou\*, Emmanuel Stratakis\**

### S1. Structural and sensing properties of the metal halide perovskite microcrystals

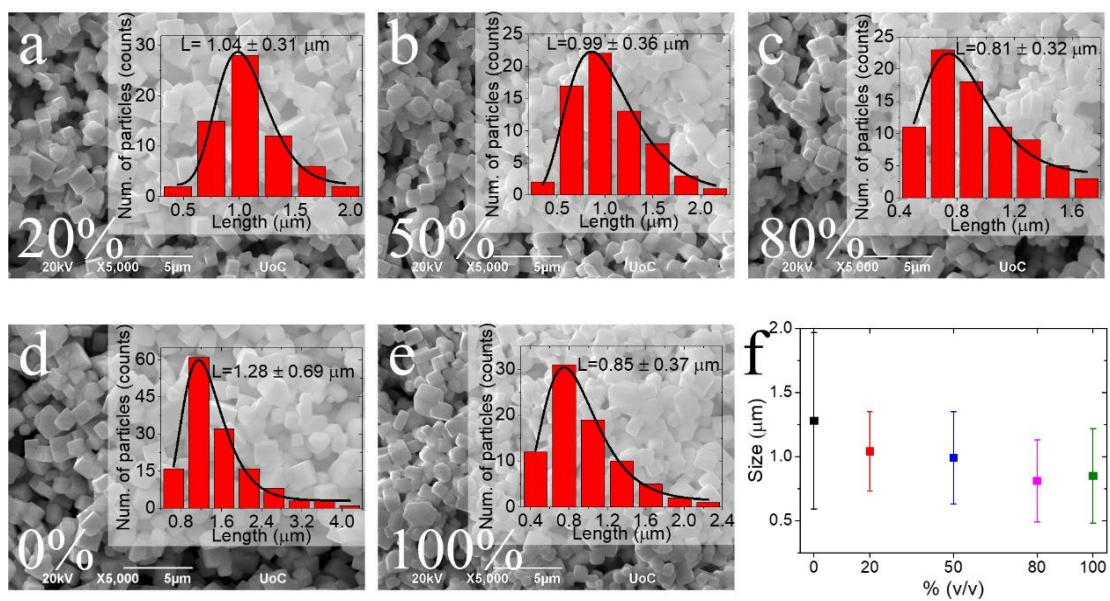

**Figure S1.** SEM images of undoped mixed halide perovskites with volume-to-volume ratios of a) 20%, b) 50%, c) 80% and the reference samples d) $CsPbBr_3$, 0%, e) $CsPbCl_3$, 100% v/v. The insets present the size distribution diagrams and the average μC size for each sample. f) Average μCs' size evolution by tuning the volume-to-volume ratio.

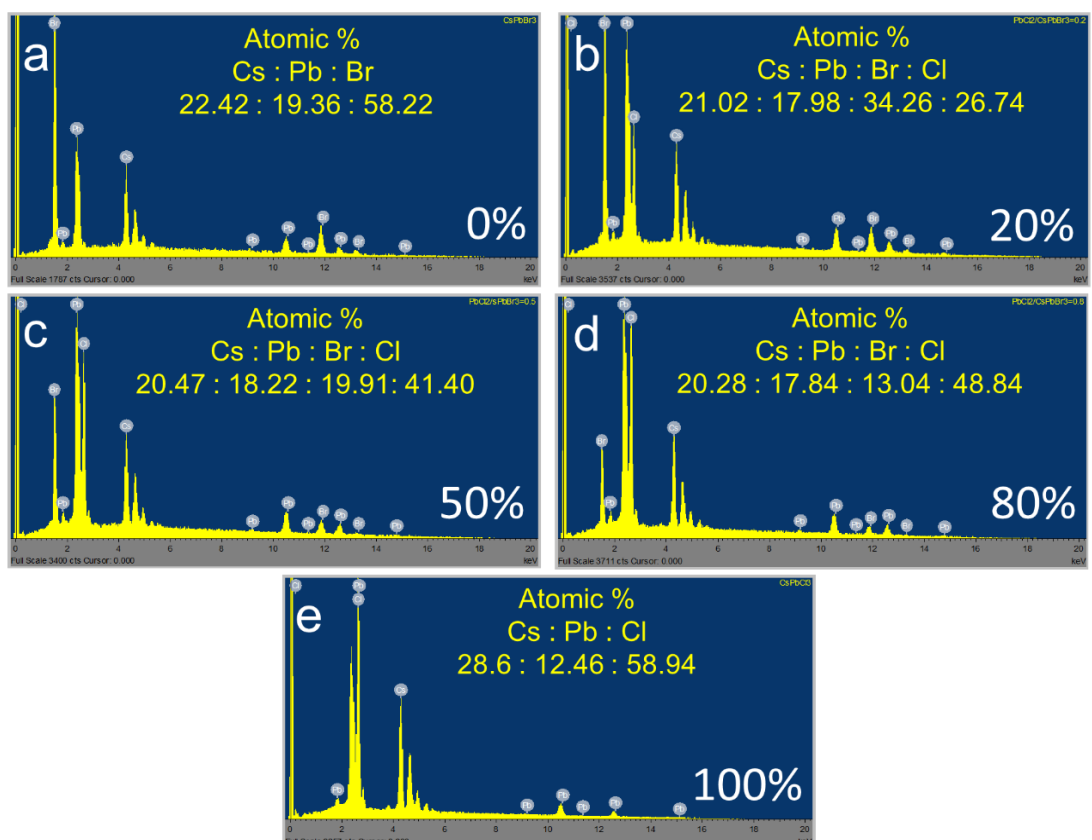

**Figure S2.** EDS spectra of the undoped $CsPbBr_{3-x}Cl_x$ perovskite μCs synthesized by varying the volume-to volume-ratio from 0% to 100%.

**Table S1.** Chemical composition of the undoped $CsPbBr_{3-x}Cl_x$ μCs as derived by EDS.

| % v/v | Chemical composition | Cs (at.%) | Pb (at.%) | Br (at.%) | Cl (at. %) |
|---|---|---|---|---|---|
| 0 | $CsPbBr_3$ | 22.42 | 19.36 | 58.22 | - |
| 20 | $CsPbBr_{1.4}Cl_{1.6}$ | 21.02 | 17.98 | 34.26 | 26.74 |
| 50 | $CsPbBr_1Cl_2$ | 20.47 | 18.22 | 19.91 | 41.40 |
| 80 | $CsPbBr_{0.08}Cl_{2.2}$ | 20.28 | 17.84 | 13.04 | 48.84 |
| 100 | $CsPbCl_3$ | 28.6 | 12.46 | - | 58.94 |

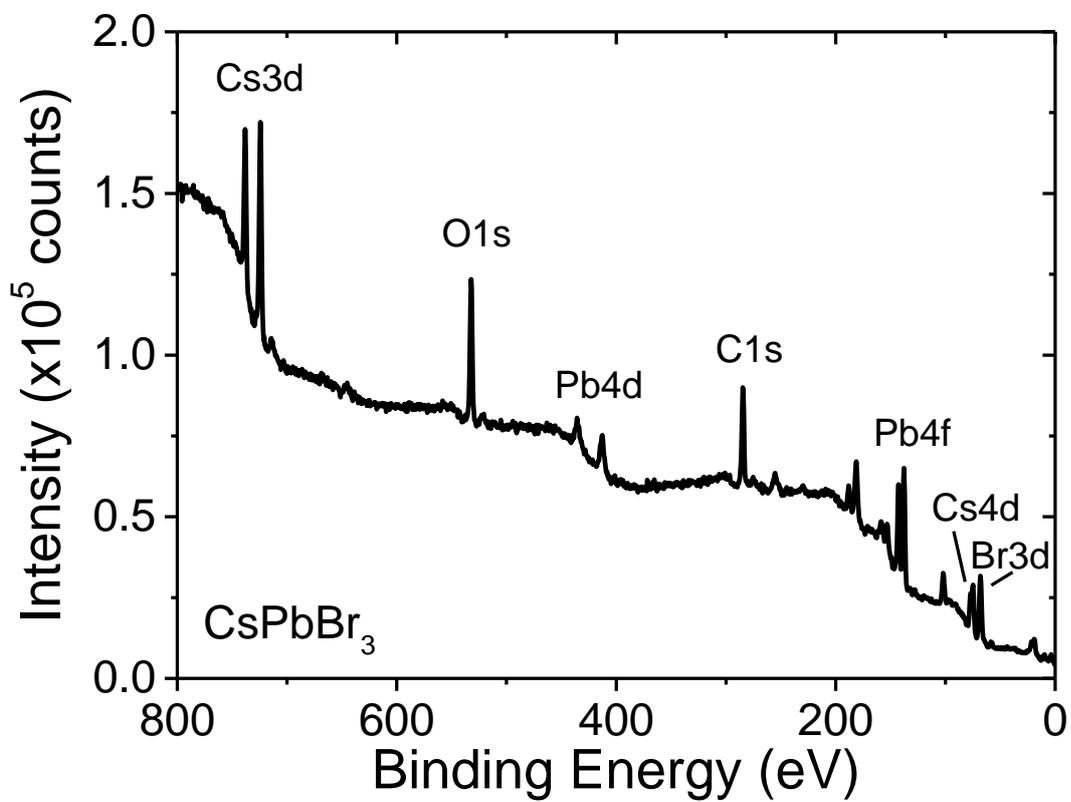

**Figure S3.** XPS survey spectrum of CsPbBr₃ μCs.

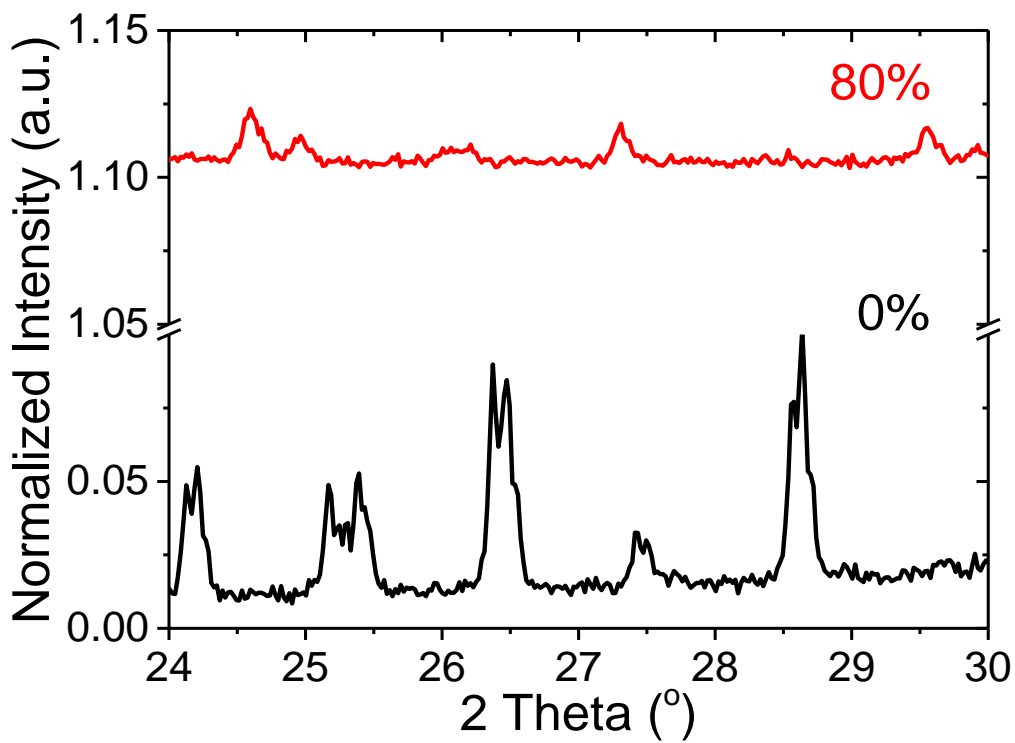

**Figure S4.** XRD patterns of 0% v/v CsPbBr$_3$ (black line) and 80% v/v undoped CsPbBr$_{3-x}$Cl$_x$ (red line) μCs.

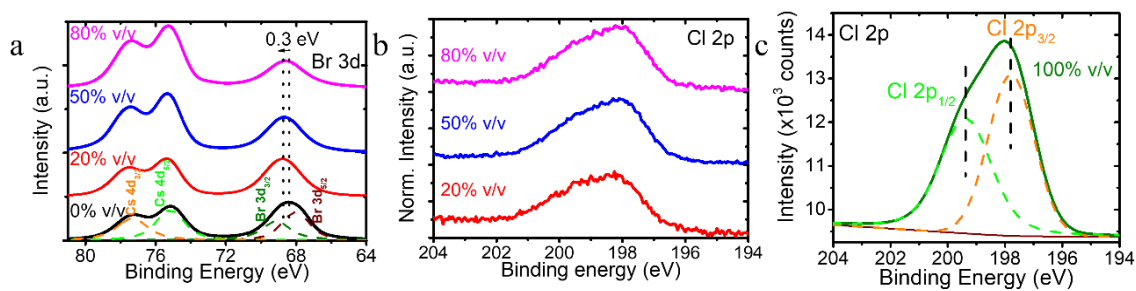

**Figure S5.** High resolution XPS spectra of a) Br 3d, b) and c) Cl 2p of 0% v/v CsPbBr$_3$ (black curve), 20% v/v undoped (red curves), 50% v/v undoped (blue curves), 80% v/v undoped (magenta curves) CsPbBr$_{3-x}$Cl$_x$ μCs and 100% v/v CsPbCl$_3$ (green curve). A thorough deconvolution of Br 3d and Cl 2p peaks of CsPbBr$_3$ and CsPbCl$_3$ respectively, revealed the typical d$_{5/2}$/d$_{3/2}$ and p$_{3/2}$/p$_{1/2}$ doublets at ~68.0 eV/~69.1 eV and ~197.8 eV/~199.4 eV for both peaks, which are associated with Pb-Br and Pb-Cl bonds ().

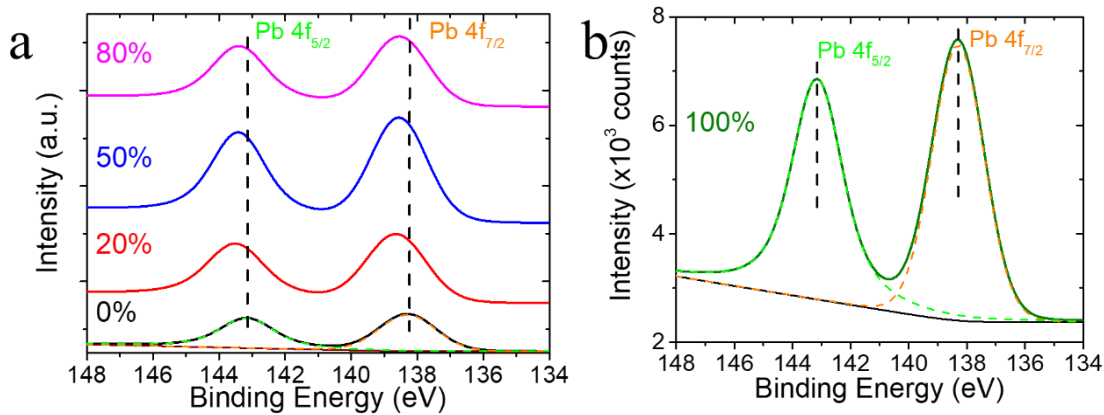

**Figure S6.** High resolution XPS spectra of Pb 4f of a) undoped $CsPbBr_{3-x}Cl_x$ µCs for different volume-to-volume ratio and reference samples, 0% $CsPbBr_3$ µCs and b) the reference sample, 100% v/v $CsPbCl_3$ µCs.

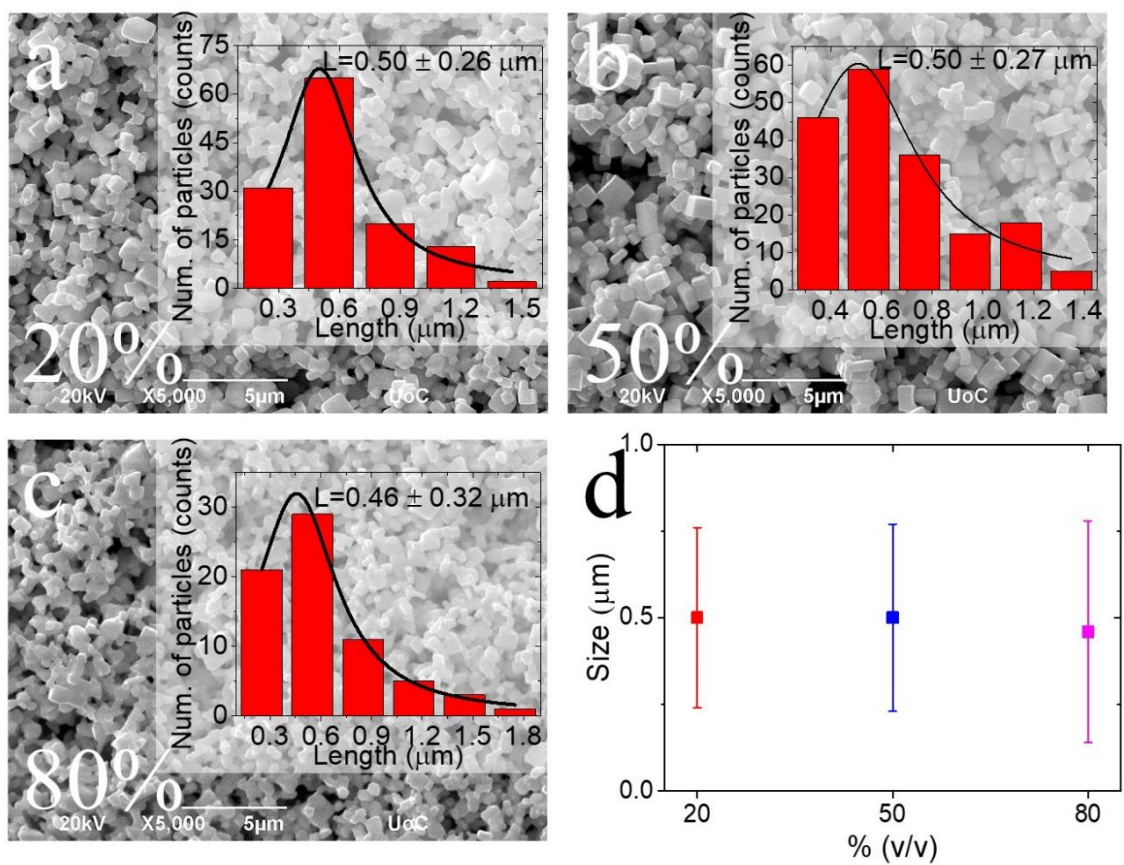

**Figure S7.** SEM images of Mn-doped $CsPbBr_{3-x}Cl_x$ µCs with volume-to-volume ratio of a) 20%, b) 50% and c) 80% v/v. The insets showed the size distribution diagrams and the average µC size for each sample. d) Average µCs' size evolution by tuning the volume-to-volume ratio.

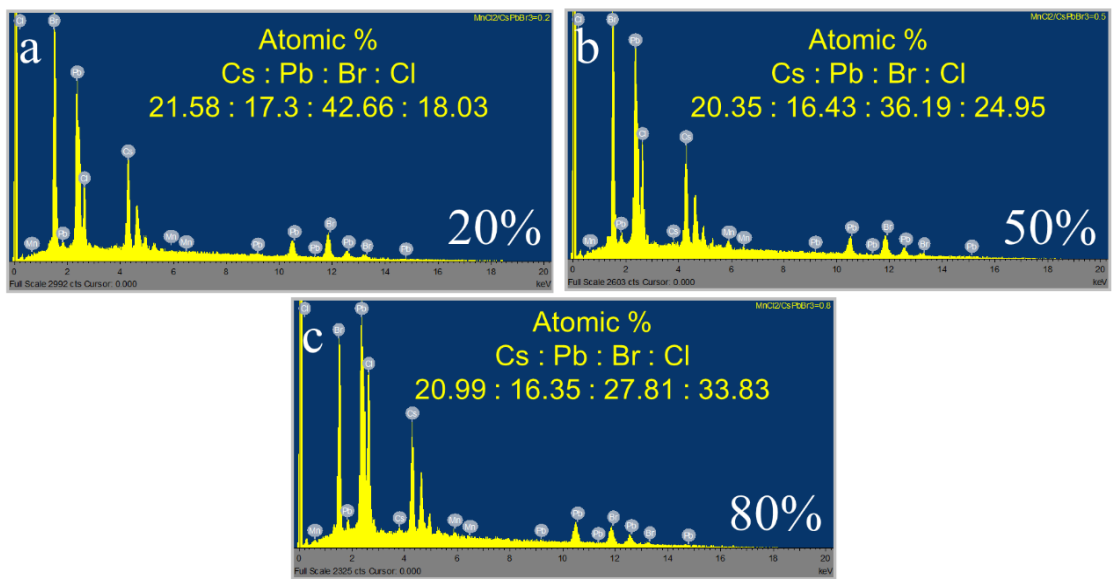

**Figure S8.** EDS spectra of Mn-doped CsPbBr$_{3-x}$Cl$_x$ μCs synthesized with a) 20%, b) 50% and c) 80% v/v.

**Table S2.** Chemical composition of the Mn-doped CsPbBr$_{3-x}$Cl$_x$ μCs as derived by EDS.

| % v/v | Cs (at.%) | Pb (at.%) | Br (at.%) | Cl (at.%) |
|---|---|---|---|---|
| 20 | 21.68 | 17.38 | 42.82 | 18.12 |
| 50 | 21.30 | 16.26 | 31.96 | 30.48 |
| 80 | 20.66 | 16.73 | 25.46 | 37.15 |

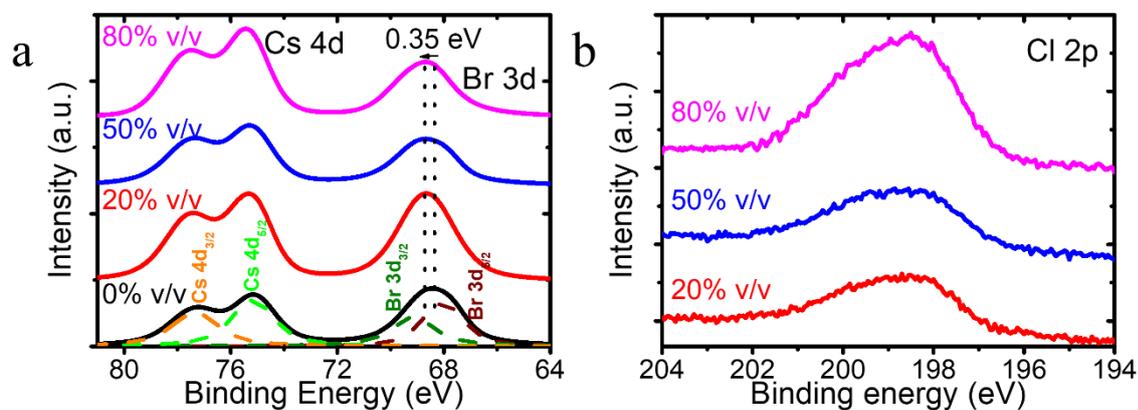

**Figure S9.** High resolution XPS spectra of a) Br 3d and b) Cl 2p of 0% v/v CsPbBr$_3$ (black curve) and Mn-doped CsPbBr$_{3-x}$Cl$_x$ μCs with varying halide ratios.

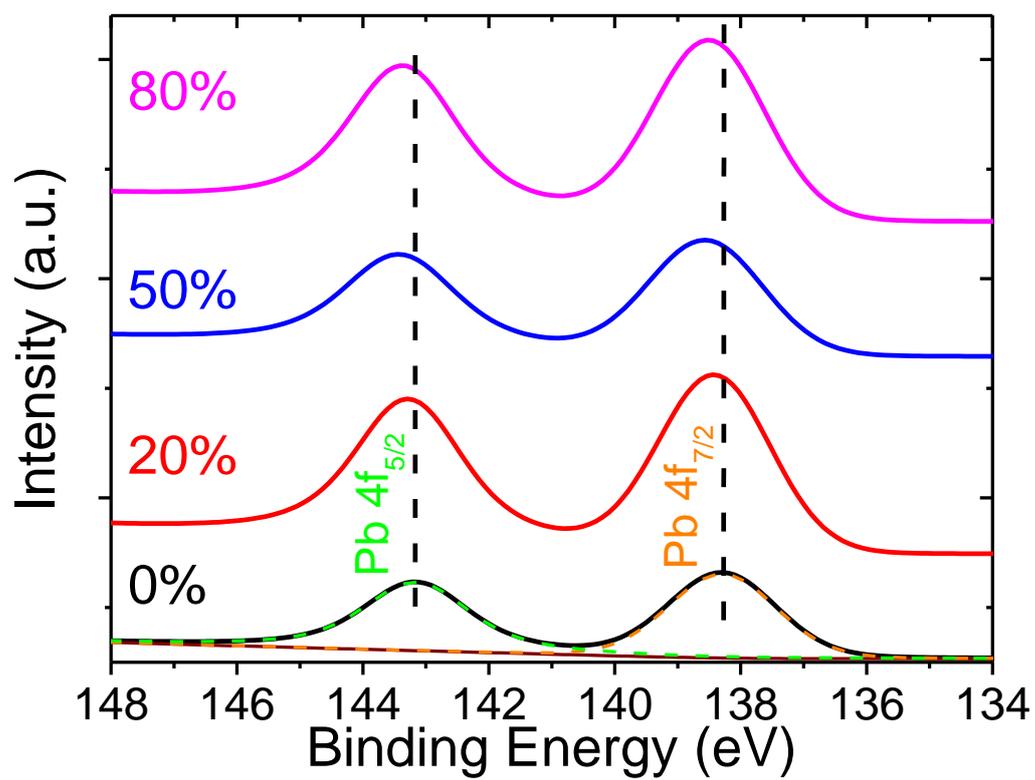

**Figure S10.** High resolution XPS spectra of Pb 4f of Mn-doped CsPbBr$_{3-x}$Cl$_x$ μCs with varying halide ratios.

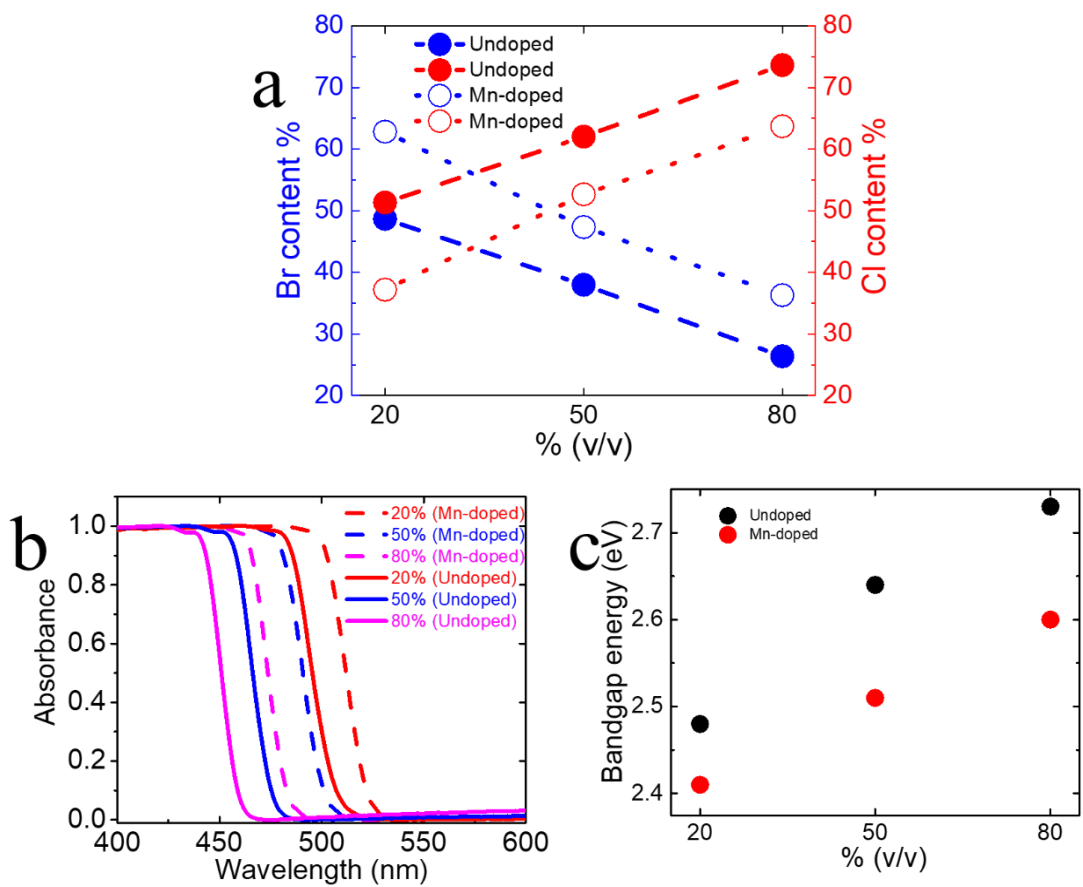

**Figure S11.** a) Halide content of undoped and Mn-doped and CsPbBr$_{3-x}$Cl$_x$ μCs as calculated by XPS survey scans. b) Absorbance spectra of undoped (compact lines) and Mn-doped (dash lines) CsPbBr$_{3-x}$Cl$_x$ μCs. c) Calculated bandgap energy of undoped (black circle) and Mn-doped (red circle) CsPbBr$_{3-x}$Cl$_x$ μCs.

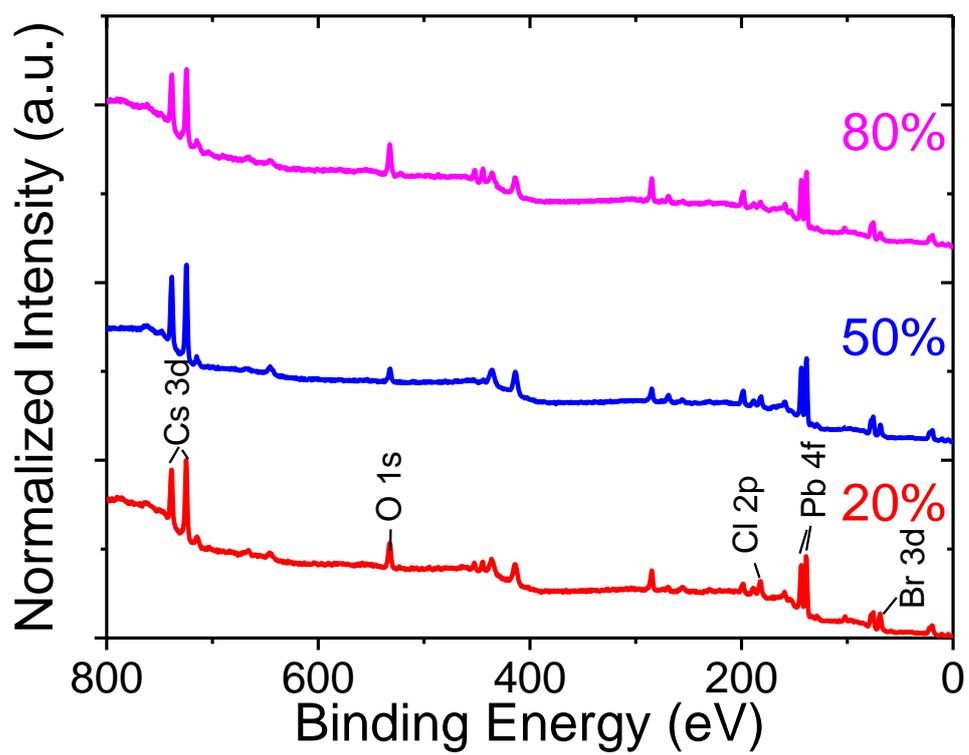

**Figure S12.** XPS survey spectra for the Mn-doped $CsPbBr_{3-x}Cl_x$ μCs by varying the v/v ratio.

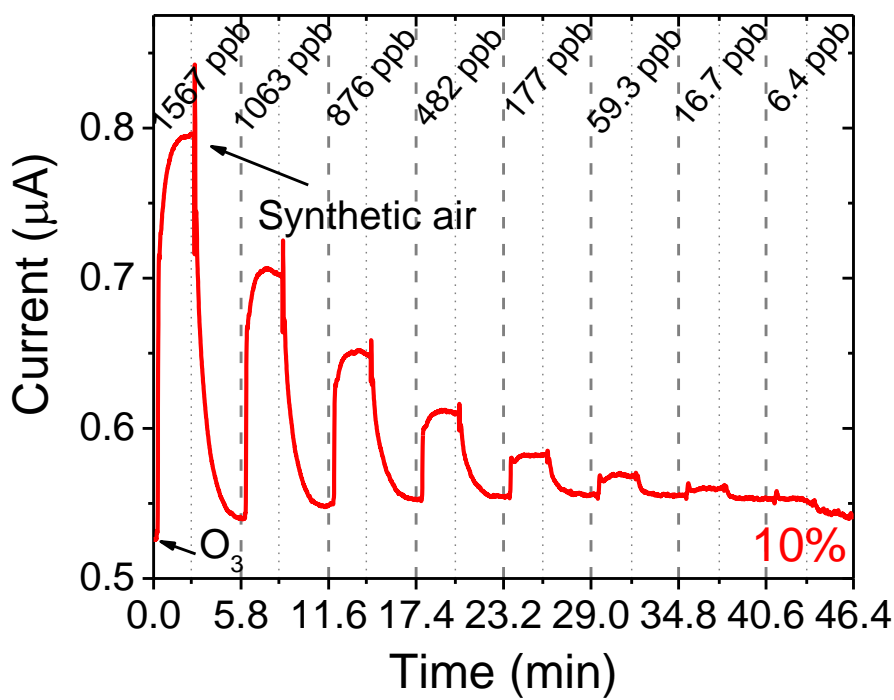

**Figure S13.** $O_3$ sensing performance of the 10% v/v undoped perovskite based-sensor.

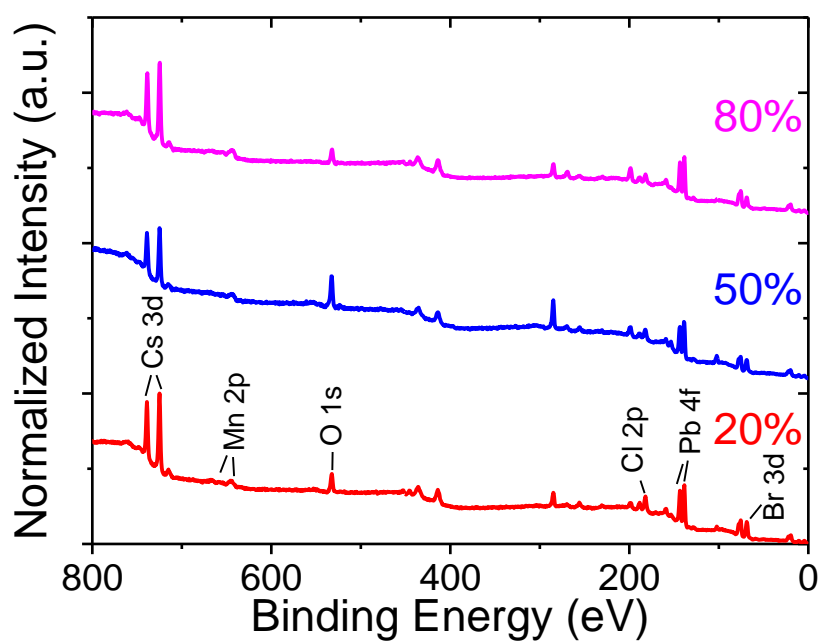

**Figure S14.** XPS survey spectra from bottom to top of 20% v/v (red line), 50% v/v (blue line) and 80% v/v (magenta line) undoped mixed halide μCs.

### S2. DFT calculations

Modelling sensing based on first-principles calculations presents inherent challenges. While real-world sensing materials exhibit diverse morphologies and are often covered with various adsorbates at ambient conditions, DFT simulations generally assume idealized scenarios In the vast majority of such simulations, an ideal surface is assumed with only one kind of adsorbates, and adsorbates are arranged in a perfect periodic overlayer. These conditions are only met under ultra-high vacuum experiments at very low temperatures with large single-crystal samples. This discrepancy, known as the "pressure gap" in surface chemistry , highlights the difference between experimental conditions and simulations.[43] Despite this gap, DFT simulations remain a valuable tool for identifying trends in surface reactivity. Experimental findings frequently confirm DFT predictions regarding the relative strength of molecular bindings, even though absolute binding energies may not align precisely.

*S2.1 Adsorption energies*

To construct a comprehensive understanding of the sensing mechanism, our investigation extended to the following model systems: (f) $CsPbBr_3$ with Cl doping, where all eight surface Br atoms of the outmost surface layer are substituted by Cl atoms, (g) tetragonal $CsPbCl_3$, and (h) tetragonal $CsPbCl_3$ with one Cl vacancy site in the outmost surface layer.

**Table S3.** Adsorption energies of $O_2$ adsorbed on the Cl-doped (001) surface of $CsPbBr_3$ and on the defect-free and defected with Cl-vacancy (001) surface of $CsPbCl_3$

| Model surface | $\Delta E_{ads}$ (eV) |
|---|---|
| 1 Cl-doped $CsPbBr_3$ | 0.73 |
| 8 Cl-doped $CsPbBr_3$ | 0.75 |
| Defect-free $CsPbCl_3$ | 0.76 |
| $CsPbCl_3$ with Cl-vacancy | -1.87 |

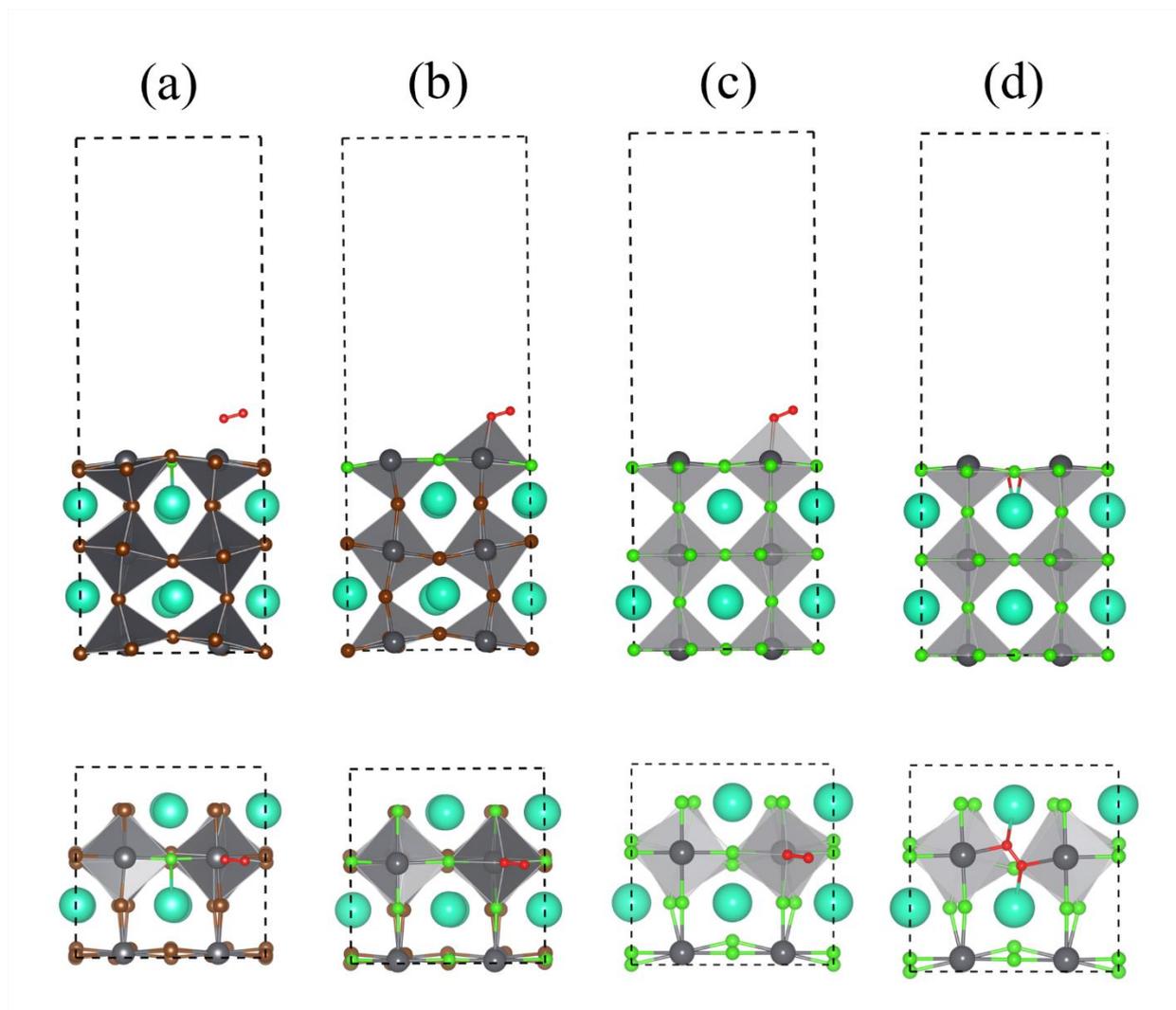

**Figure S15.** DFT-optimized configurations of the (001) oriented model surfaces; a) 1Cl-doped and b) 8Cl-doped CsPbBr$_3$ model surface, viewed along b (upper line) and c (bottom line) axis. c) Defect-free CsPbCl$_3$ and d) with Cl-vacancy model surface. Cs, Pb, Br, Cl, and O atoms are illustrated as cyan, gray, brown, green, and red atoms, respectively. Dashed lines indicate the periodic boundaries of the simulation supercell.

*S2.2 Electronic Properties*

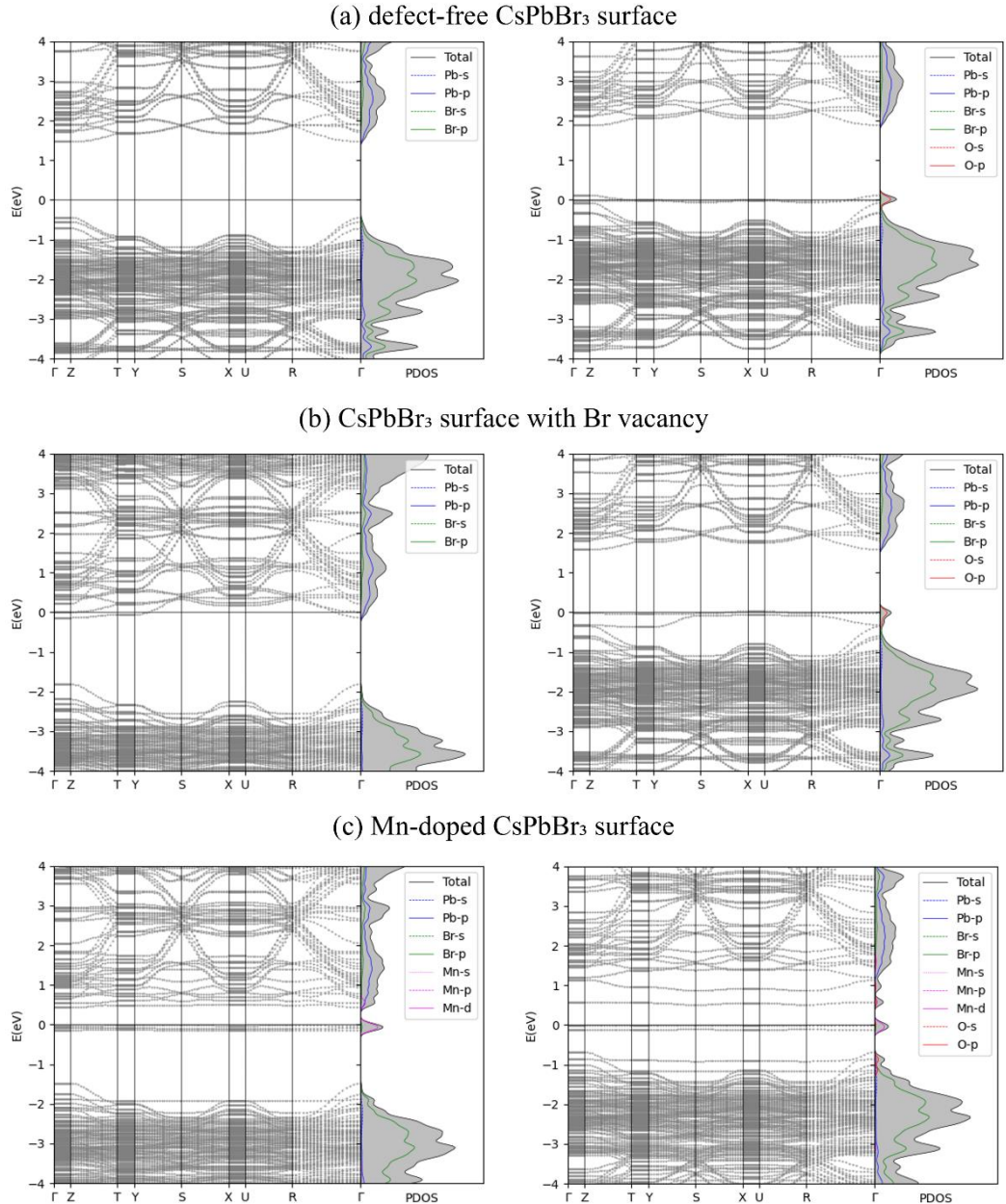

**Figure S16.** Band structures (BS) and partial density of states (PDOS) of the (001) oriented CsPbBr$_3$ model surfaces without a) defects, b) with Br vacancy, and c) Mn-doped. For each system, the left panel corresponds to the model surfaces without oxygen, while the right panel refers to the surfaces with O$_2$ adsorbed.

As it can be seen in Figure S16 (a), the valence band of defect-free surface is mainly occupied by the p orbitals of Br atoms, while the conduction band derives its contributions from the p orbitals of Pb atoms. When molecular oxygen is adsorbed on the defect-free surface, its p orbitals introduce new shallow acceptor states at the top of the valence band. Comparing the defect-free surface to the one with a Br vacancy (Figure S16b), new shallow states that provide electrons emerge at the bottom of the conduction band once a Br vacancy is introduced. However, upon oxygen adsorption on the Br vacancy, these shallow donor states are withdrawn, and new shallow acceptor states emerge at the top of the valence band, deriving their contributions from the p orbitals of oxygen atoms.

A rather interesting case emerges with the introduction of Mn to the lattice (Figure S16c) and was highlighted in the main text. Here, not only new donor states are observed in the conduction band, but also localized states from the d orbitals of Mn appear within the bandgap. Upon adsorption of oxygen on our Mn-doped model surface, the shallow donor states are quite withdrawn, and new acceptor states appear at the top of the valence band. The d-states of Mn remain localized at the same relative position within the gap, while the oxygen states exhibit a non-negligible density located both at the valence band and close to the conduction band minimum.

The band structure is altered also in the case of a Pb vacancy (Figure S17a) compared to the defect free case, with both the conduction and valence band appearing higher in energy. Oxygen adsorption in the Pb-vacancy introduces shallow acceptor states at the top of the valence band, which derive their contributions from the p orbitals of O and Br atoms.

For the halide substitution case, where one surface Br atom is substituted by one Cl atom (Figure S17b), we observe minor changes in the electronic structure. All the qualitative BS and PDOS characteristics remain similar to the defect-free case, as Cl states are located deep within the BS, far away from the CBM and VBM. When more Br atoms are substituted with Cl (Figure S16c), the contribution of Cl states in the VB becomes more significant, up to the case of $CsPbCl_3$ (Figure S18a and b). Interestingly, the Fermi level shows a tendency to shift towards that of an n-doped system (Figure S19-22).

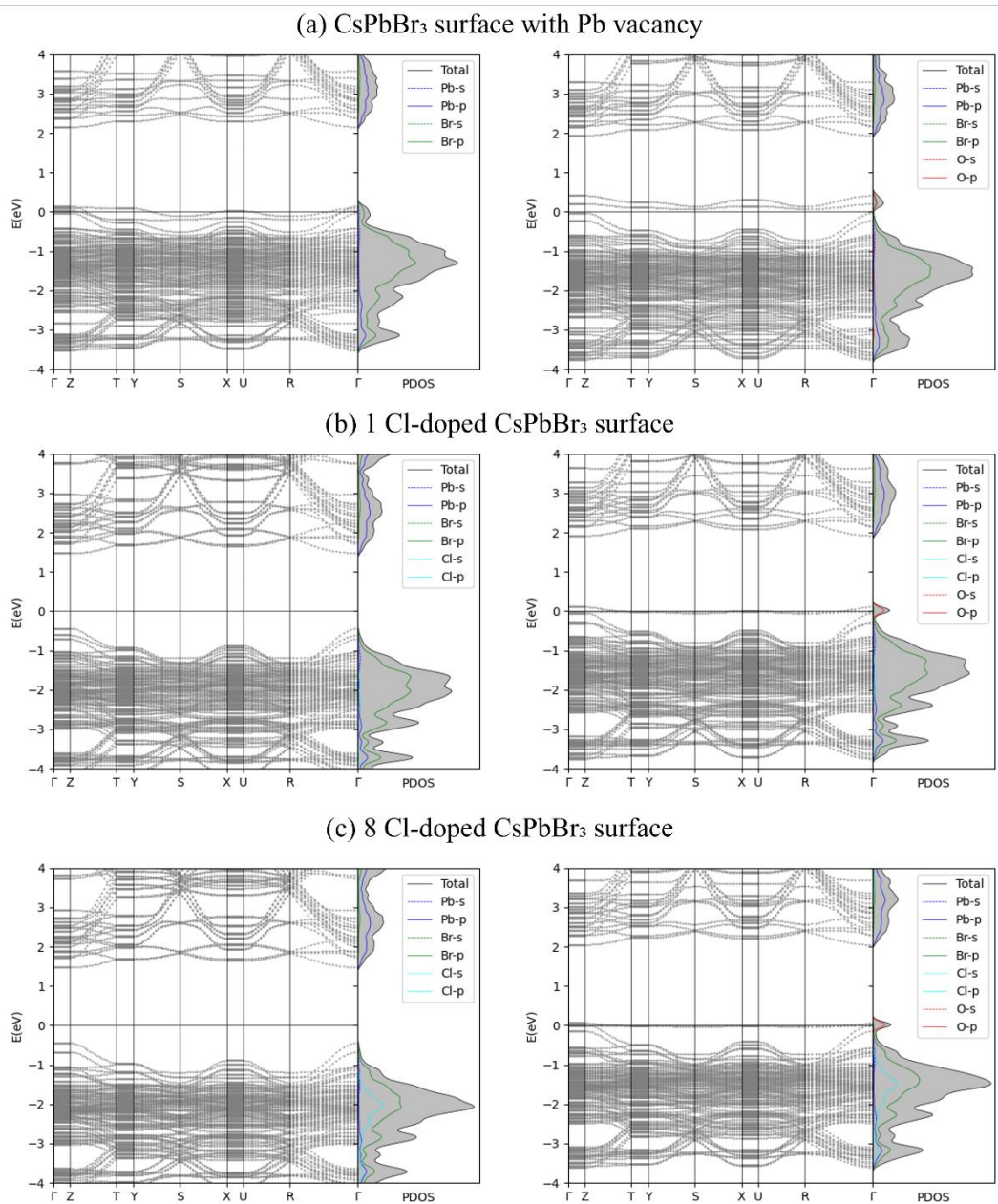

**Figure S17.** Band structures (BS) and partial density of states (PDOS) of the (001) oriented CsPbBr$_3$ model surfaces; with a) Pb vacancy, b) 1Cl-doped and c) 8Cl-doped. For each system, the left panel corresponds to the model surfaces without oxygen, while the right panel refers to the surfaces with O$_2$ adsorbed.

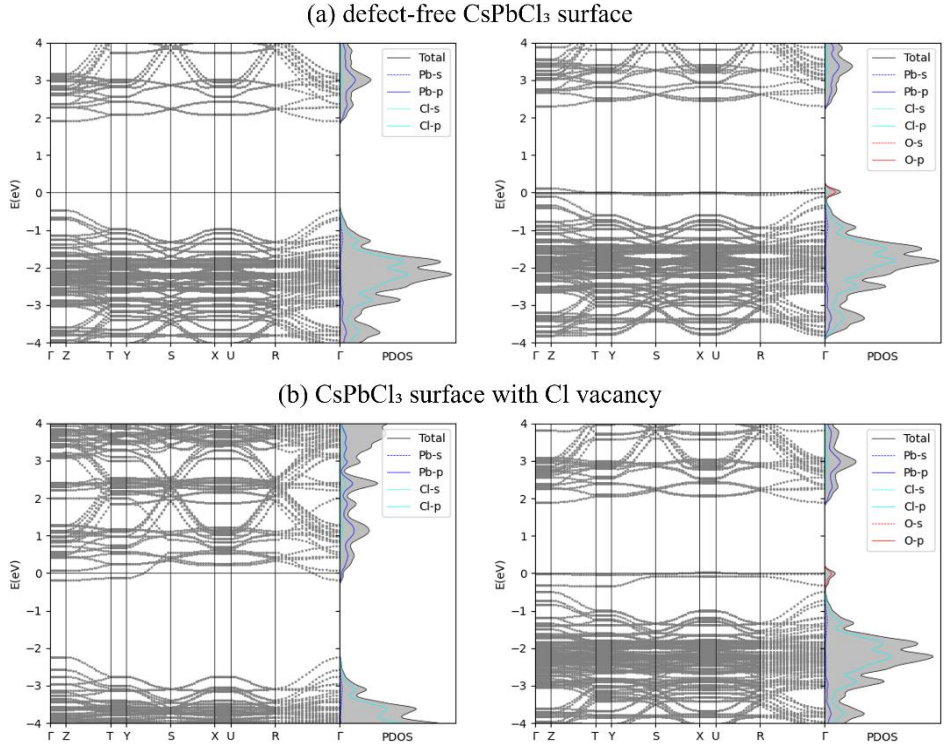

**Figure S18.** Band structures (BS) and partial density of states (PDOS) a) of the defect-free (001) oriented $CsPbCl_3$ model surface and b) with Cl-vacancy. For each system, the left panel corresponds to the model surfaces without oxygen, while the right panel refers to the surfaces with $O_2$ adsorbed.

*S2.3 Fermi shift*

In order to have a proper indication for the doping type of our model systems, we calculate the relative difference in their Fermi levels. Since DFT is well-known for accurately characterizing the DOS peaks below the Fermi level, we used the lowest energy peaks in the range -28 to -18 eV, which are identical in all systems (Figure S18). These peaks were found to have contributions only from the s orbitals of the Cs atoms in the case of the surfaces without $O_2$ adsorbed and from Cs-s and O-s orbitals upon adsorption of molecular oxygen; orbitals that do not contribute to bond formation. Using the peaks shown in Figure S18a and the peaks from -22 to -19 eV shown in Figure S18b we shift the peaks of all systems (colored lines) towards the peak of the defect-free slab (denoted in black), a process demonstrated in Figure S19.

Indicatively, upon adsorption of molecular oxygen in the defect-free $CsPbBr_3$ model surface (Figure S20a), states from the p-orbitals of oxygen emerge and the fermi level sinks 0.4 eV, suggesting a p-type behavior. Once a Br vacancy is introduced (Figure S20b), the fermi level rises 1.6 eV and "enters" the VB, indicating an n-type behavior. Once oxygen fills the Br vacancies, the n-type behavior diminishes as holes are provided by oxygen to the system, switching to a p-type behavior. In the case of oxygen adsorption on the Cl substituted $CsPbBr_3$ model surface (Figure S21b and c), when more Cl atoms are substituting the Br atoms of the uppermost layer of the surface, the Fermi level shows a tendency to move towards that of an n-doped system.

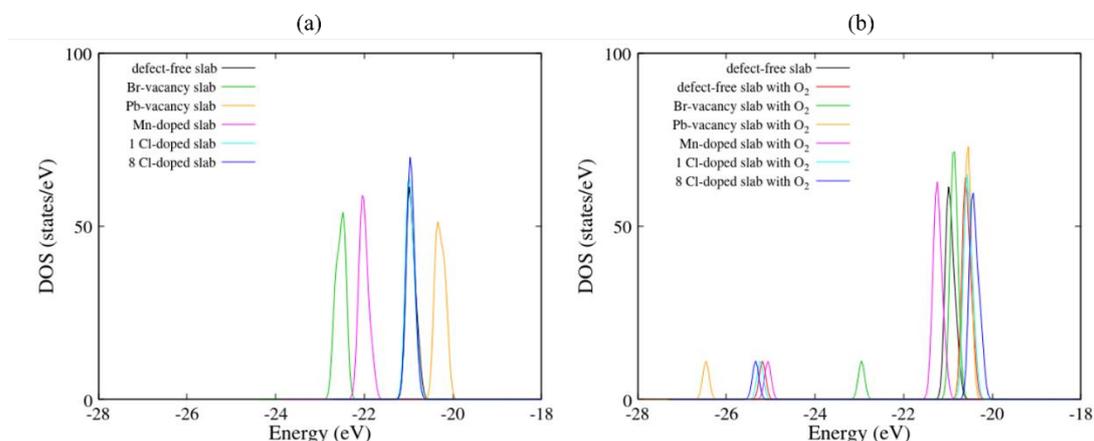

**Figure S19.** The lowest energy peaks of the DOS of all model $CsPbBr_3$ surfaces without $O_2$ adsorbed in (a) and upon adsorption of $O_2$ in (b).

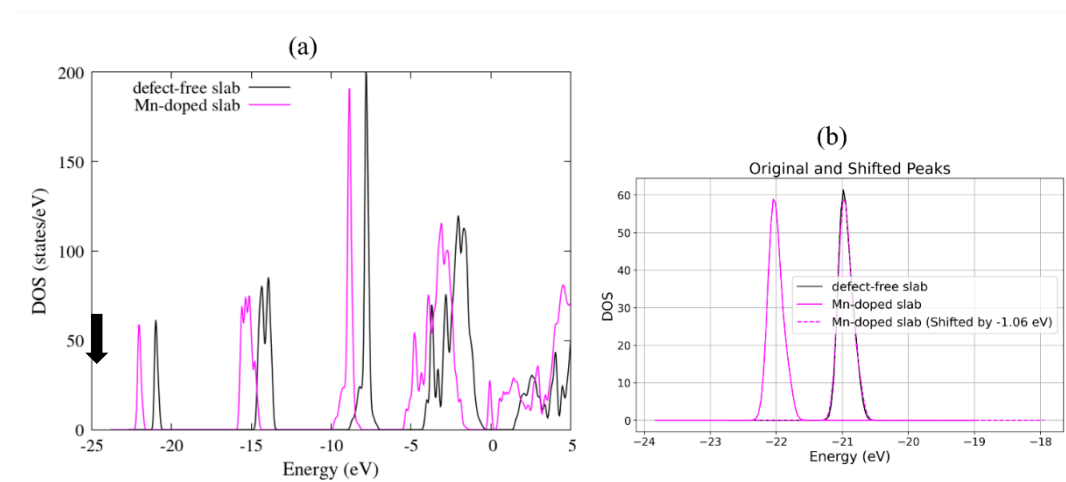

**Figure S20.** a) Total DOS of the defect-free and the Mn-doped $CsPbBr_3$ slab in black and magenta, respectively. b) The lowest energy identical peaks (denoted in (a)) used to calculate the shift of the Fermi level.

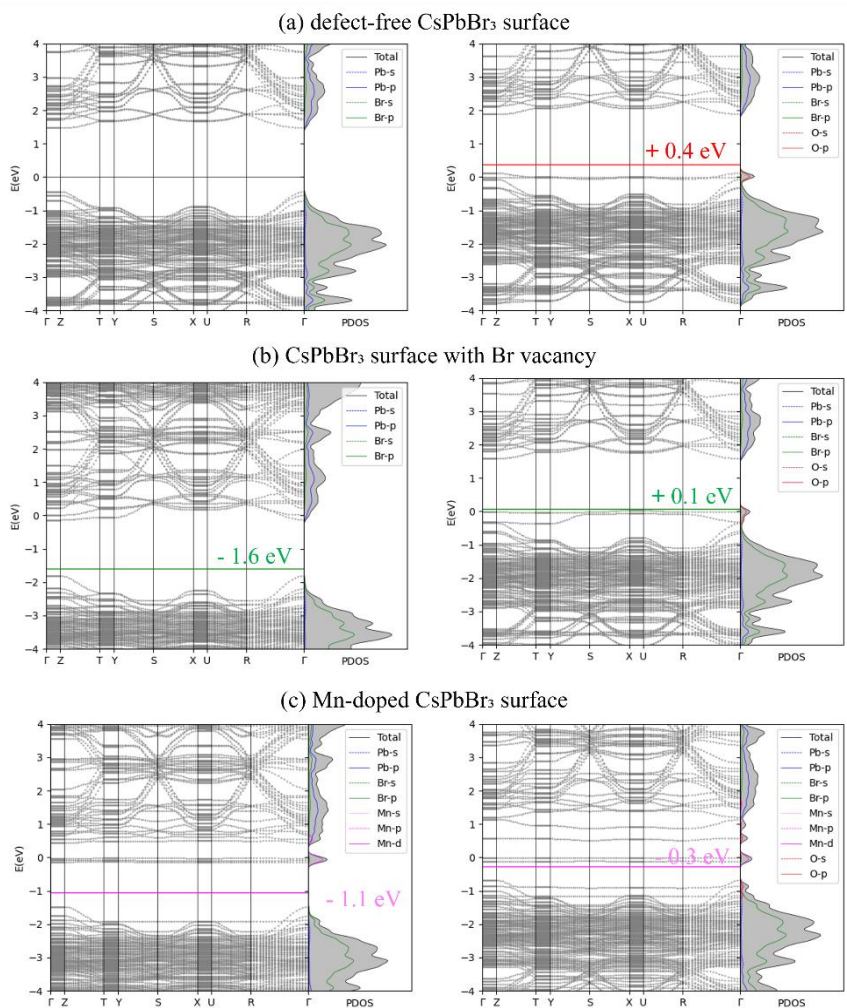

**Figure S21.** Band structures (BS) and partial density of states (PDOS) of the (001) oriented $CsPbBr_3$ model surfaces a) without defects, b) with Br vacancy, and c) Mn-doped. For each system, the left panel corresponds to the model surfaces without oxygen, while the right panel refers to the surfaces with $O_2$ adsorbed. Colored lines represent the fermi level shift of the defect-free system in each case.

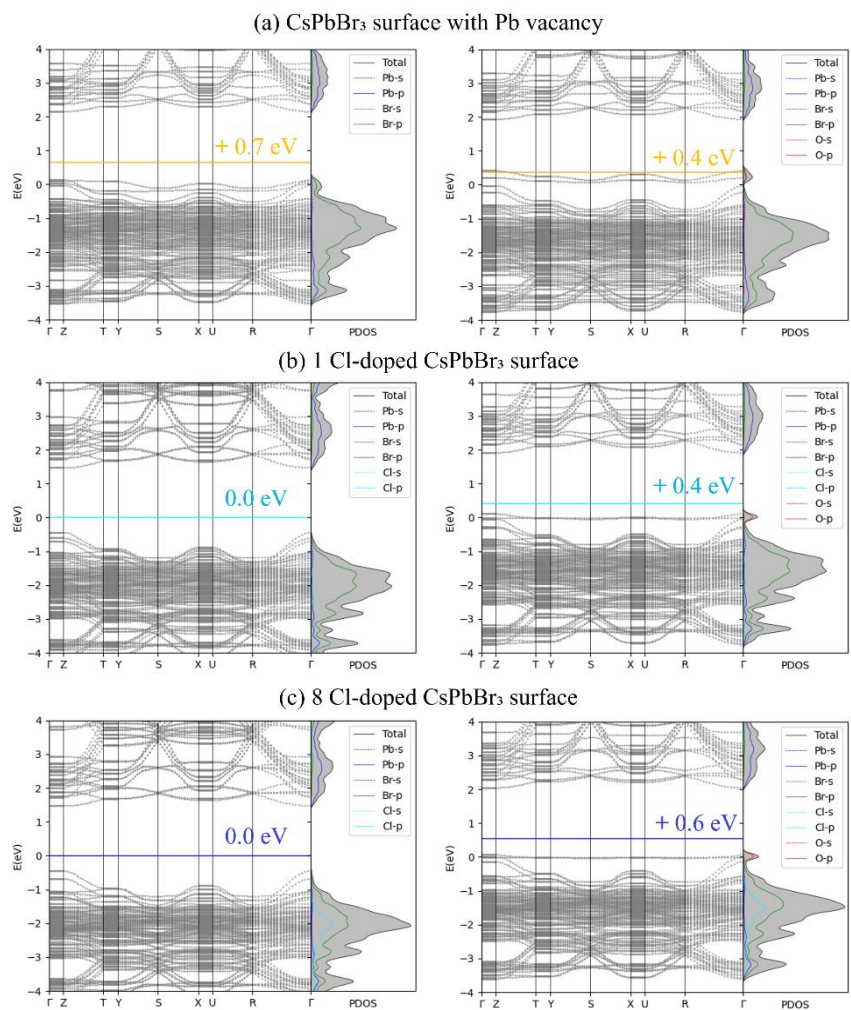

**Figure S22.** Band structures (BS) and partial density of states (PDOS) of the (001) oriented CsPbBr$_3$ model surface with a) Pb vacancy, b) 1Cl-doped, and c) 8Cl-doped. For each system, the left panel corresponds to the model surfaces without oxygen, while the right panel refers to the surfaces with O$_2$ adsorbed. Colored lines represent the fermi level shift of the defect-free system in each case.

*S3. Aging study and sensing properties*

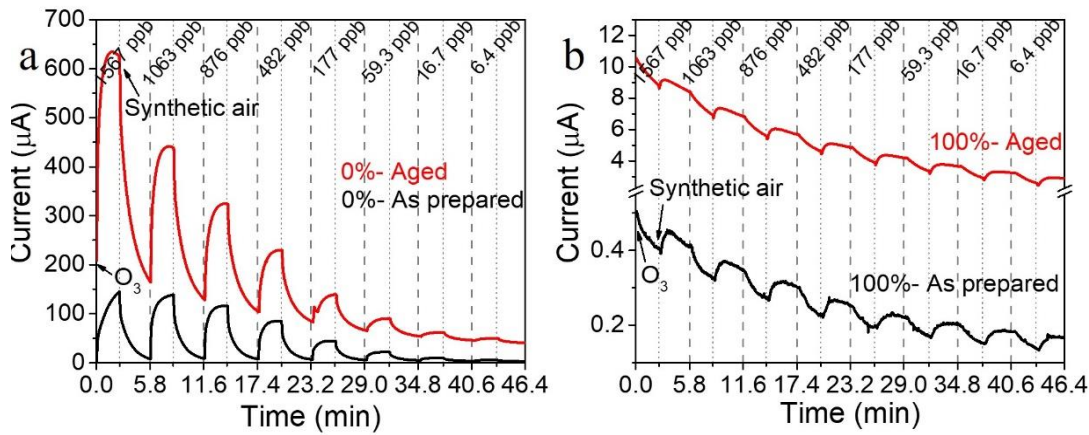

**S23.** O$_3$ sensing performance of the as-prepared (black curves) and aged for a month (red curves) reference samples, a) 0% v/v CsPbBr$_3$ and b) 100% v/v CsPbCl$_3$ µCs.

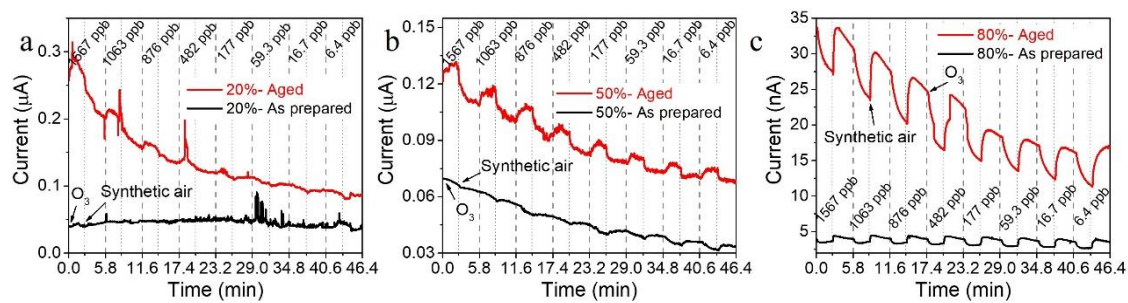

**Figure S24.** O$_3$ sensing performance of a) 20%, b) 50% and c) 80% v/v, as prepared (black lines) and aged (red lines) undoped perovskite-based sensors.

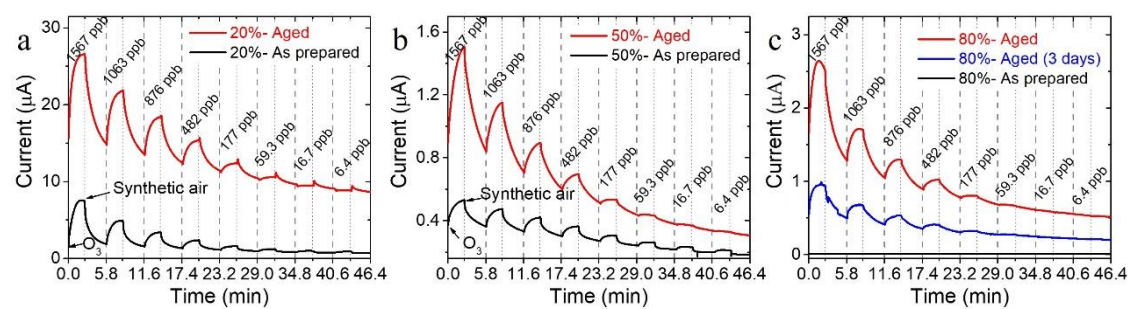

**Figure S25.** O$_3$ sensing performance of a) 20%, b) 50% and c) 80% v/v, as prepared (black lines) and aged (red lines) Mn-doped perovskite-based sensors.

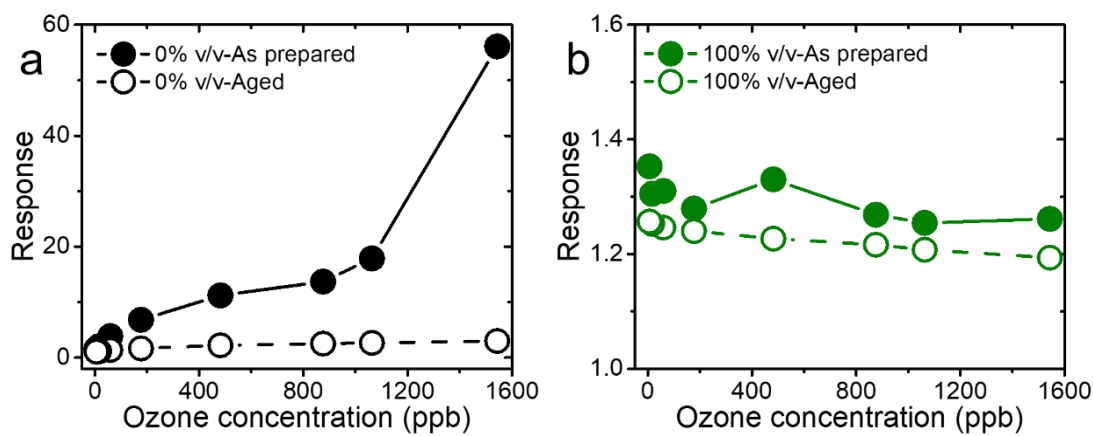

**Figure S26.** Calculated response as a function of $O_3$ concentration of a) 0% $CsPbBr_3$ and b) 100% v/v $CsPbCl_3$ perovskite-based sensors as prepared (filled circle) and after a few weeks (hollow circle).

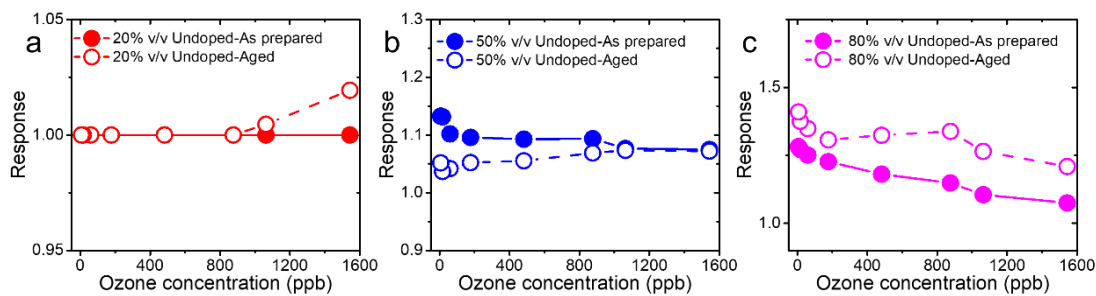

**Figure S27.** Calculated response as a function of $O_3$ concentration of a) 20%, b) 50% and c) 80% v/v as prepared (filled circle) and aged for a month (hollow circle) undoped perovskite-based sensors.

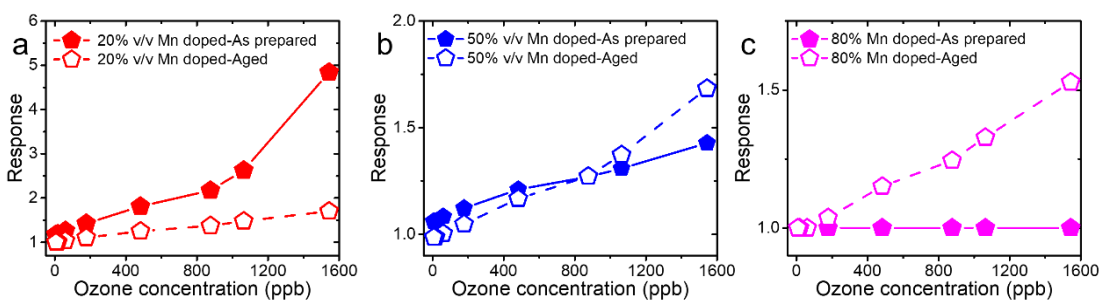

**Figure S28.** Calculated response as a function of $O_3$ concentration of a) 20%, b) 50% and c) 80% v/v, as prepared (filled pentagon) and aged (hollow pentagon), Mn-doped perovskite-based sensors.

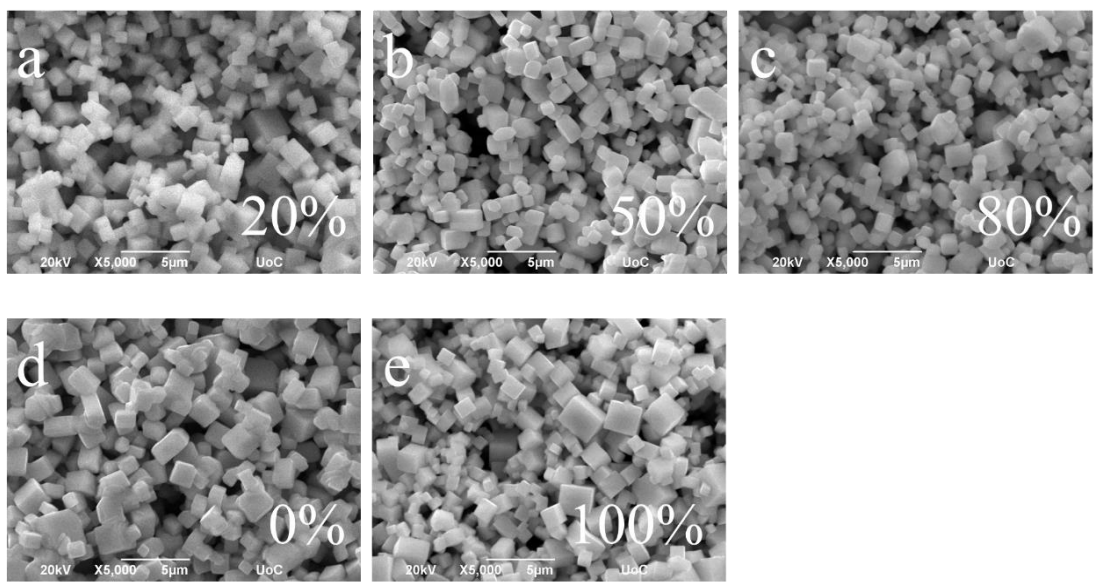

**Figure S29.** SEM images of a) 20%, b) 50%, c) 80% v/v undoped mixed halide perovskite μCs and the reference samples d) 0% v/v $CsPbBr_3$ and e) 100% v/v $CsPbCl_3$ μCs after exposure to air for 50 days.

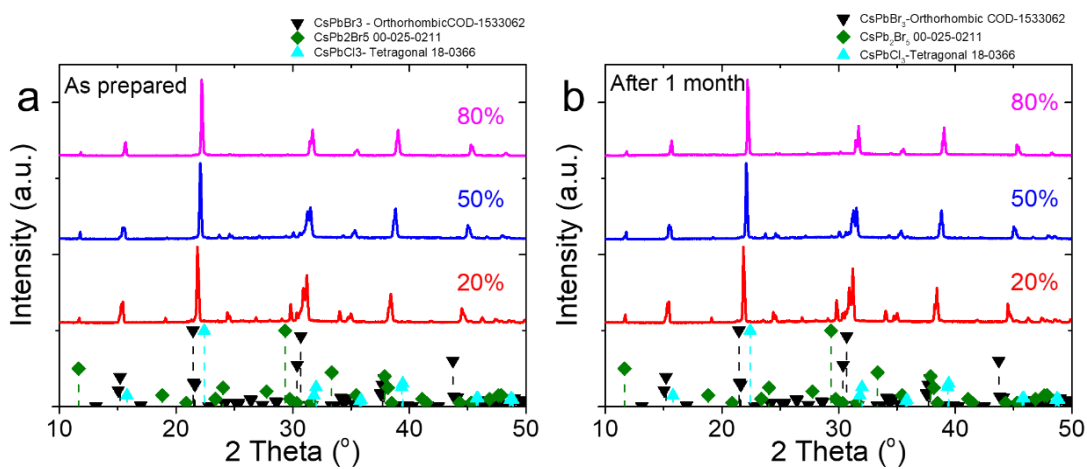

**Figure S30.** XRD patterns of 20% (red line), 50% (blue line) and 80% v/v (magenta line) undoped mixed halide perovskites a) as prepared, b) aged for one month and c) aged for one year.

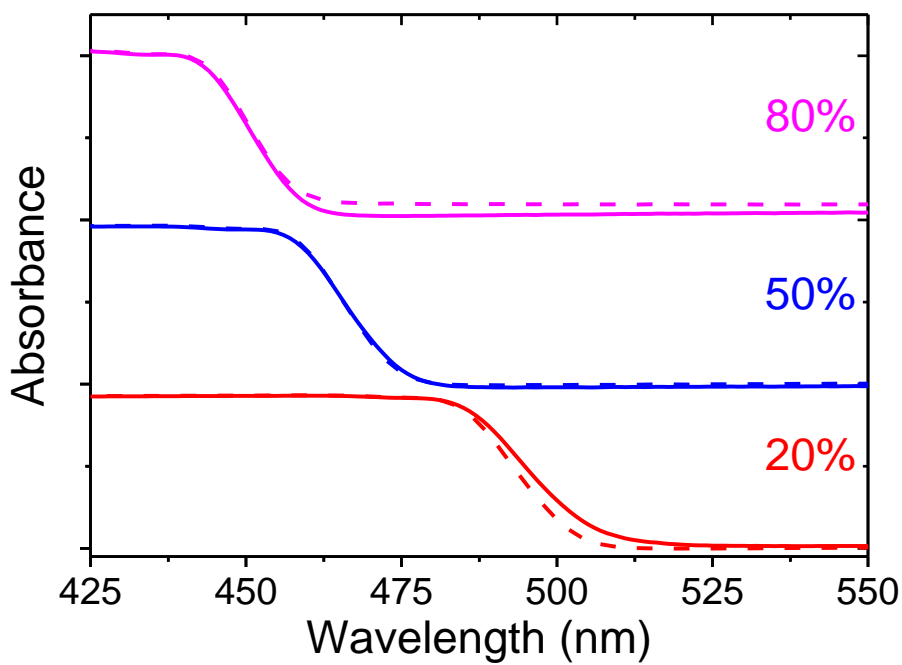

**Figure S31.** Absorbance spectra of the as prepared (solid lines) and aged for one month (dashed lines) undoped mixed halide μCs by varying the v/v ratio.

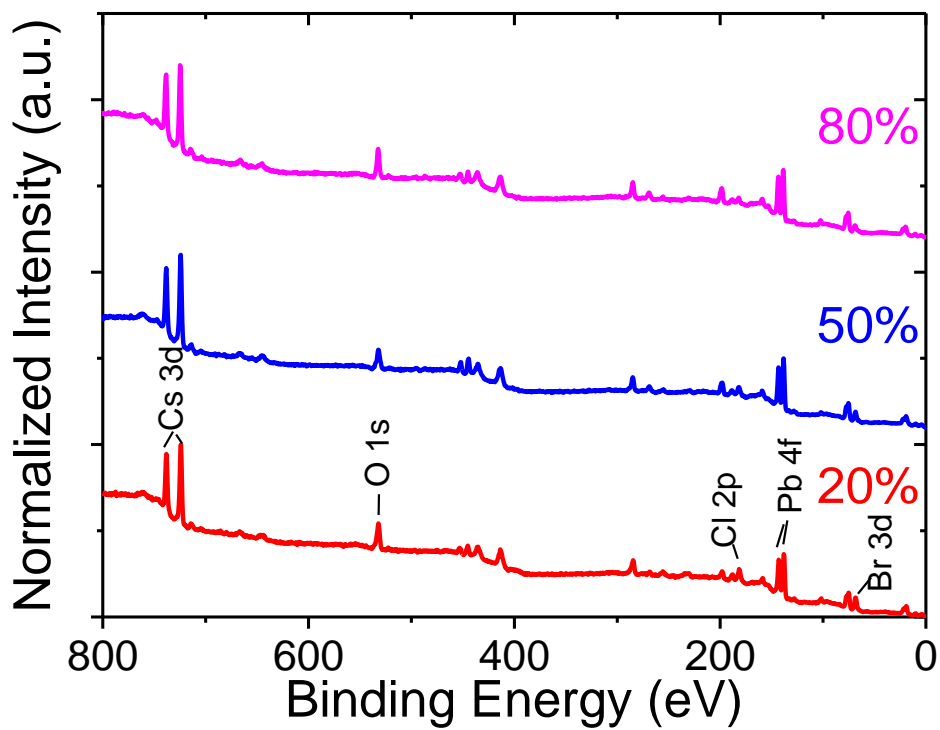

**Figure S32.** XPS survey spectra of the aged undoped mixed halide μCs by varying the v/v ratio.

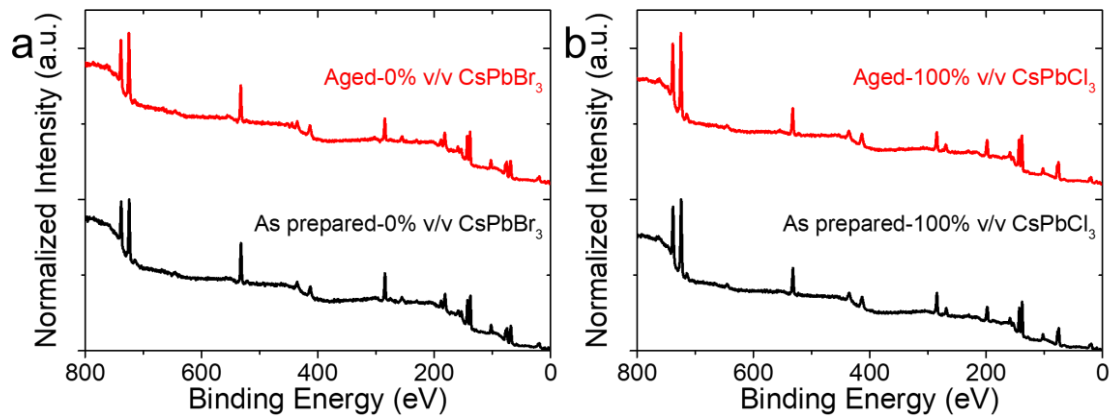

**Figure S33.** XPS survey spectra of the as prepared (black line) and aged for a month exposed to ambient conditions (red line) of the a) 0% v/v $CsPbBr_3$ and b) 100% v/v $CsPbCl_3$ μCs.

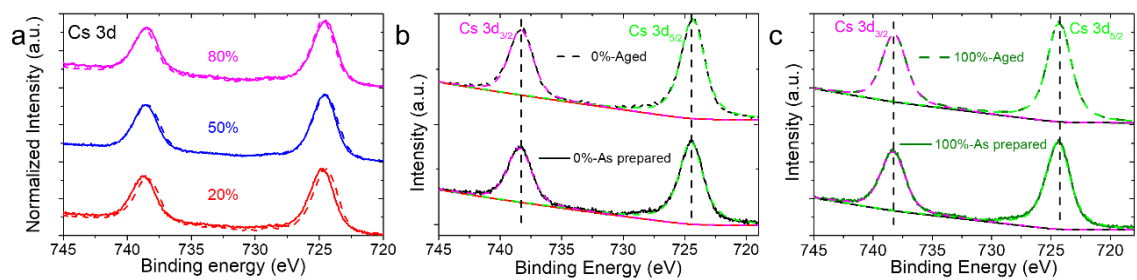

**Figure S34.** High resolution XPS spectra of Cs 3d of the as prepared (solid lines) and aged for a month to ambient conditions (dash lines) of the a) undoped mixed halide perovskite μCs, b) 0% v/v $CsPbBr_3$ and c) 100% v/v $CsPbCl_3$ μCs.

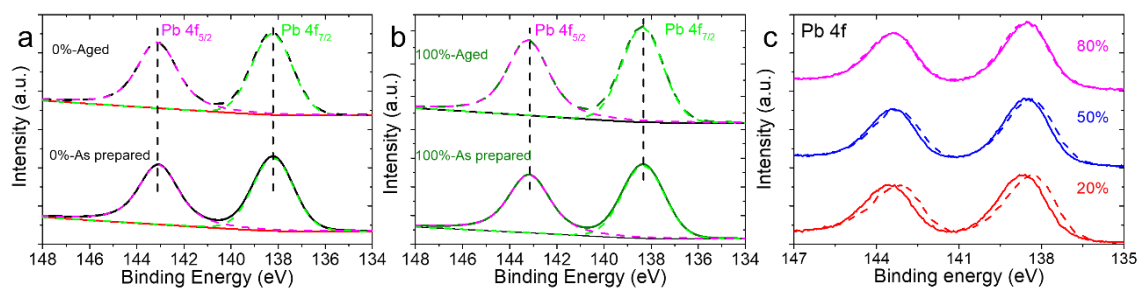

**Figure S35.** High resolution XPS spectra of Pb 4f of the as prepared (solid lines) and aged for a month to ambient conditions (dash lines) of the a) 0% v/v $CsPbBr_3$, b) 100% v/v $CsPbCl_3$ and c) undoped mixed halide perovskite, varying the v/v.

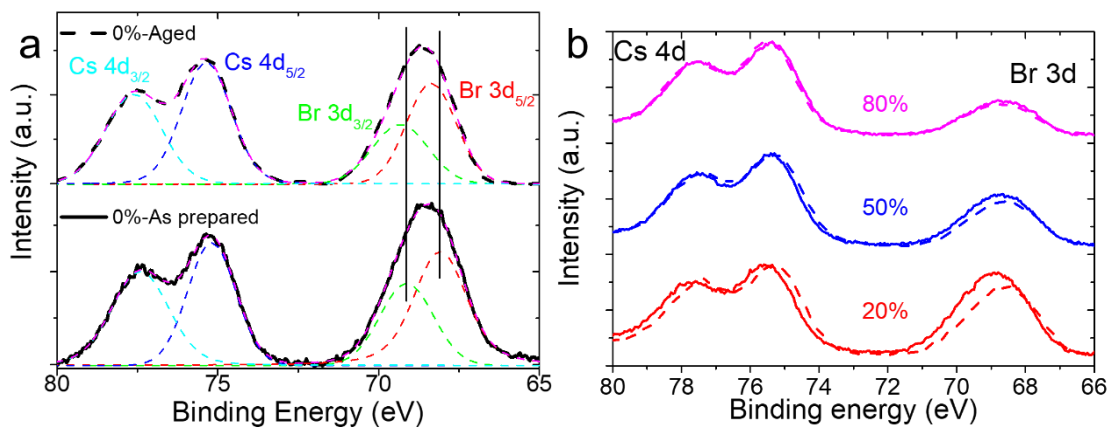

**Figure S36.** High resolution XPS spectra of Br 3d of the as prepared (solid lines) and aged for a month to ambient conditions (dash lines) of the a) 0% CsPbbr₃ and b) undoped mixed halide perovskite, varying the v/v, μCs.

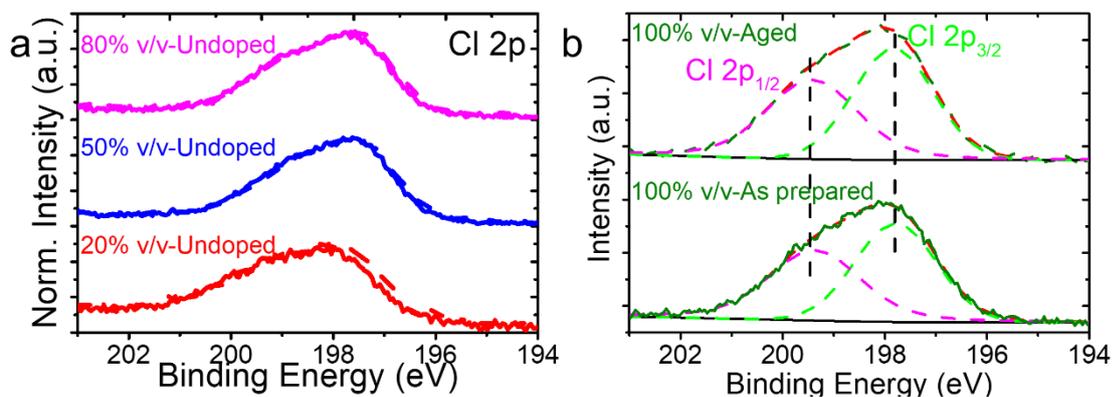

**Figure S37.** High resolution XPS spectra of Cl 2p of the as prepared (solid lines) and aged for a month to ambient conditions (dash lines) of the a) undoped mixed halide perovskite varying the % v/v and b) 100% CsPbCl₃ μCs.

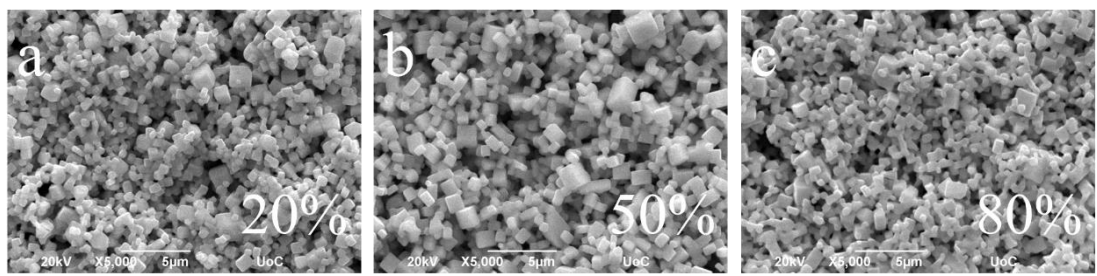

**Figure S38.** SEM images of the aged Mn-doped mixed halide perovskite μCs exposed to ambient conditions.

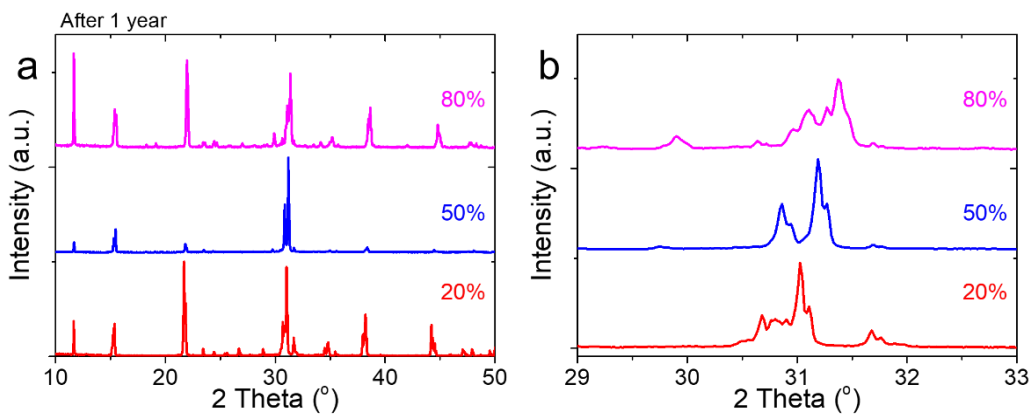

**Figure S39.** XRD patterns of Mn-doped perovskite μCs aged under ambient conditions for one year, with varying v/v ratios. XRD patterns for the 2 theta range a) from 10 to 50° and b) zooms in on the range from 29 to 33º.

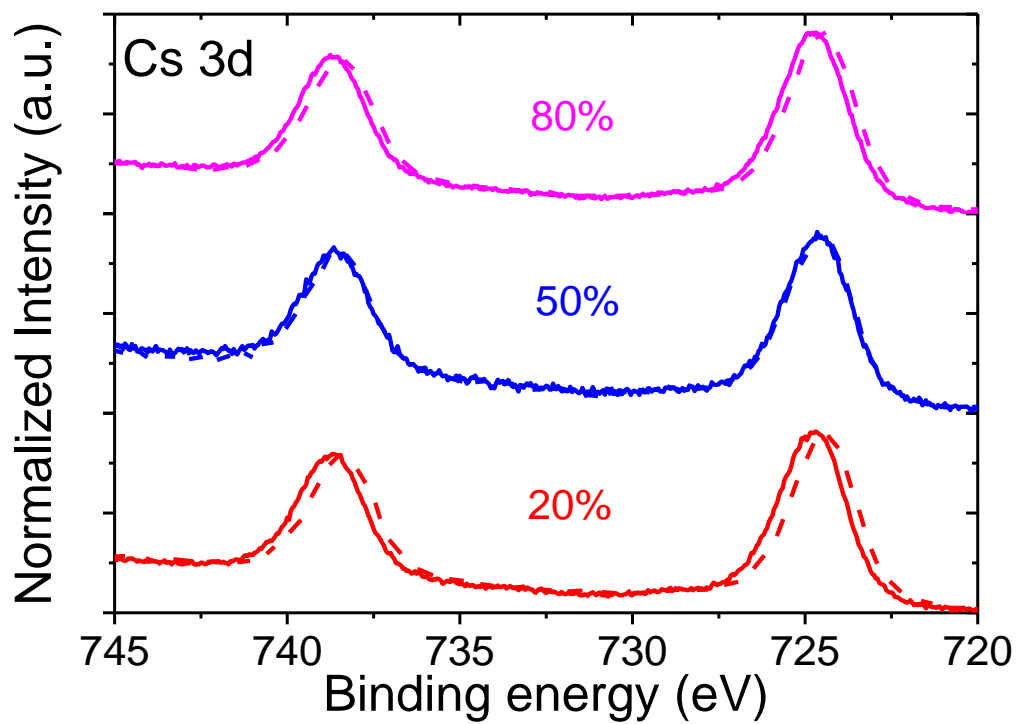

**Figure S40.** High resolution XPS spectra of Cs 3d of the as prepared (solid lines) and aged for a month to ambient conditions (dash lines) Mn-doped mixed halide perovskite μCs varying the v/v ratio.

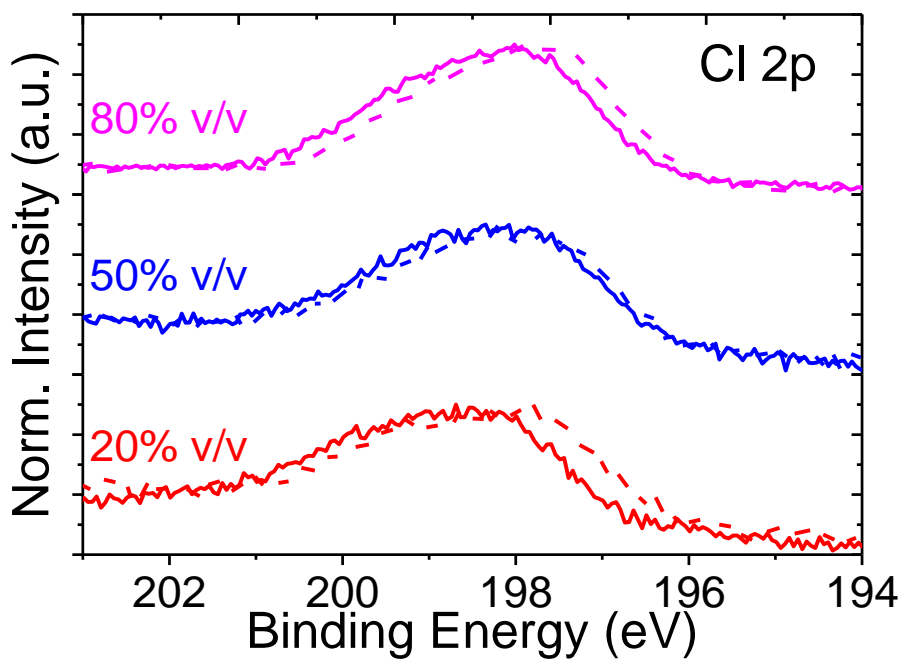

**Figure S41.** High resolution XPS spectra of Cl 2p of the as prepared (solid lines) and aged for a month to ambient conditions (dash lines) Mn-doped mixed halide perovskite μCs varying the v/v ratio.

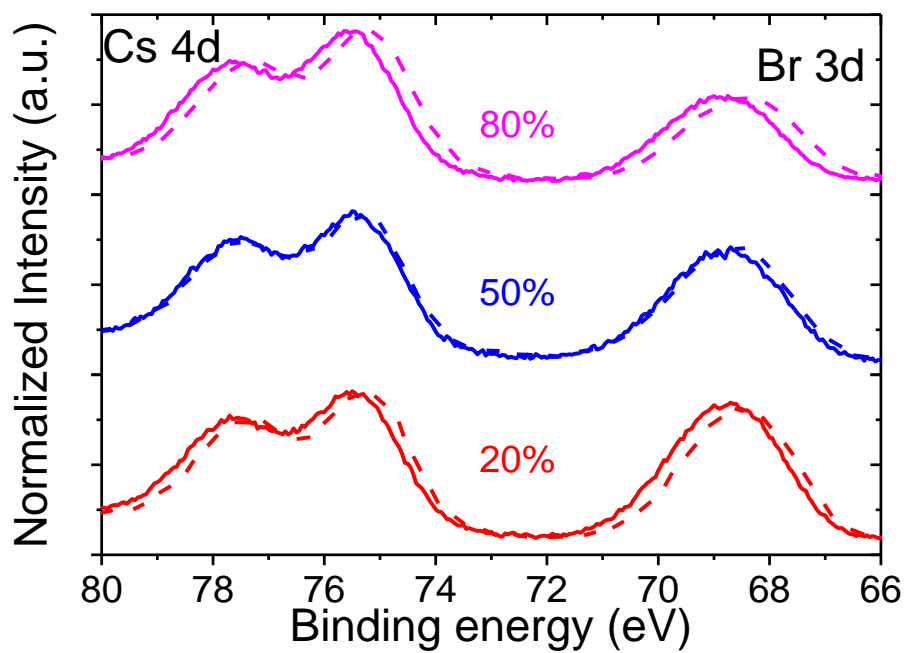

**Figure S42.** High resolution XPS spectra of Br 3d of the as prepared (solid lines) and aged for a month to ambient conditions (dash lines) Mn-doped mixed halide perovskite μCs varying the v/v ratio.

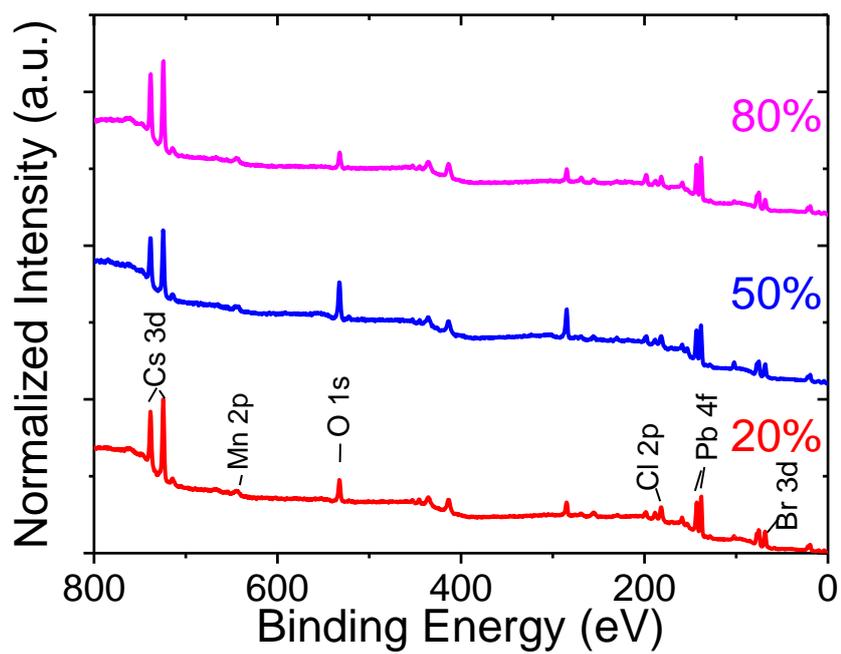

**Figure S43.** XPS survey spectra of Mn-doped mixed halide μCs by varying the v/v ratio.